\documentclass[journal]{IEEEtran}
\usepackage{cuted} 
\usepackage{lipsum} 

\usepackage{enumitem}
\usepackage[compress]{cite}

\usepackage{color,soul}
\usepackage{xcolor}
\definecolor{Mygreen}{RGB}{0,60,0}
\usepackage{amssymb}
\usepackage{amsmath}
\usepackage{amsfonts}
\usepackage{amsthm}
\usepackage{gensymb}
\usepackage{pifont}

{}

  {}
  {}

\usepackage{comment}
\usepackage{graphicx}
\usepackage{caption}
\usepackage{subcaption}
\usepackage{epstopdf}
\graphicspath{ {./fig/} } 
\usepackage{booktabs}
\usepackage{multirow}
\usepackage{multicol}
\usepackage{enumitem}
\usepackage{float}
\usepackage{textcomp}
\usepackage{upgreek}
\usepackage{lineno}
\usepackage{algpseudocode} 
\usepackage{algorithm}
\usepackage{comment}

\usepackage{array}
\begin{document}
\title{An Integer Linear Programming Approach for Maximum Power Extraction from Solar PV Plants under Partial Shading}
\author{\IEEEauthorblockN{Ushnik Chakrabarti\IEEEauthorrefmark{1}, Binoy Kumar Karmakar\IEEEauthorrefmark{2},~\IEEEmembership{Senior Member,~IEEE}\\
}
\IEEEauthorblockA{\IEEEauthorrefmark{1}University of Calcutta, Kolkata}
\IEEEauthorblockA{\IEEEauthorrefmark{2}University of Calcutta, Kolkata}
}
 
\maketitle
\vspace{-250ex}
\begin{abstract}
Partial shading in solar photovoltaic (SPV) plants, particularly in urban environments, is a common challenge caused by nearby trees, buildings, or other fixed obstructions, leading to a significant reduction in overall system efficiency. Dynamic and static PV array reconfiguration strategies are widely regarded as effective approaches for mitigating the adverse effects of partial shading. However, Dynamic Array Reconfiguration (DAR) is rarely adopted in practical systems due to high switching complexity and substantial computational requirements. In contrast, Static Array Reconfiguration (SAR) does not require complex switching arrangements or additional computational resources, making it more suitable for real-world implementation.
However, SAR is a one-time configuration and cannot adapt to dynamically changing shading conditions. Existing SAR techniques rearrange PV modules based on assumed shading regions rather than the actual shading pattern, which limits their effectiveness under practical, time-varying conditions.
In this work, an SAR technique is proposed that explicitly considers the actual shading pattern on the PV array. The proposed approach accounts for shading caused by nearby fixed obstructions that varies throughout the day as well as across different seasons. 
The performance of the proposed technique was evaluated by comparing it with existing methods considering a PV array with a square matrix, and a small-scale laboratory prototype of non-square matrix was developed to demonstrate its practical applicability in real-world scenarios. It has been observed that the method consistently delivers an optimal power output for both software simulation and practical experiment compared to other available techniques. 

\end{abstract}

\section{Introduction}\label{sec:intro}
\textcolor{black}{
Partial shading is an unavoidable and critical challenge in solar photovoltaic (SPV) plants, particularly in urban and semi-urban environments where surrounding trees, buildings, utility poles, and other fixed obstructions frequently obstruct solar irradiance. The resulting nonuniform illumination across the PV array leads to current mismatch among modules, increased power losses, and the appearance of multiple local maxima in the power-voltage characteristics, thereby significantly reducing the overall energy yield and operational efficiency of the system. Consequently, effective mitigation of partial shading effects has become an important research focus in PV system design and operation. Several advanced MPPT techniques have been proposed to overcome these challenges \cite{bollipo2020hybrid,ahmad2019power, 10316572}. However, their performance remains limited under complex shading conditions.}

\textcolor{black}{
Among the various mitigation techniques, PV array reconfiguration has emerged as a promising and widely investigated solution. This approach aims to redistribute the impact of shading across the array by altering the electrical interconnections among PV modules, thereby improving current sharing and enhancing power extraction. Reconfiguration strategies are broadly classified into dynamic and static methods. Dynamic Array Reconfiguration (DAR) continuously adapts the array configuration in response to changing shading conditions and can achieve near optimal performance under highly variable irradiance. A wide range of Dynamic Array Reconfiguration (DAR) strategies have been reported in the literature, with most implementations relying on metaheuristic optimization algorithms inspired by natural phenomena. Genetic algorithm (GA) based approaches are capable of identifying optimal interconnections among PV modules \cite{deshkar2015solar}, but they often suffer from slow convergence. Particle swarm optimization (PSO) based reconfiguration methods demonstrate faster convergence characteristics \cite{babu2017particle}. To improve the balance between exploration and exploitation, Fathy \textit{et al.} introduced the grasshopper optimization algorithm (GOA) in \cite{fathy2018recent}, which outperforms both GA and PSO in this regard. The Harris Hawks Optimizer (HHO) \cite{yousri2020optimal} employs an adaptive two phase search mechanism and offers simpler parameter tuning compared with GOA. Grey wolf optimization (GWO) based techniques \cite{stonier2022novel} emulate the hierarchical hunting behavior of wolf packs to achieve an effective trade off between exploration and exploitation.
}

\textcolor{black}{
Further improvements have been reported using the butterfly optimization algorithm (BOA) \cite{fathy2020butterfly}, which enhances power extraction over GWO through fragrance based decision rules, and the flower pollination algorithm (FPA) \cite{ram2022new}, which combines global and local pollination strategies. More recently, white shark optimization (WSO), inspired by predator prey dynamics, has been applied to DAR in \cite{kadhim2024optimal}. In addition, a current injection based DAR technique was proposed in \cite{karmakar2021current}, which improves power output without explicitly reconfiguring the array.
}

\textcolor{black}{
Despite their effectiveness, most DAR techniques involve high computational complexity, parameter sensitivity, and premature convergence, along with complex switch matrices and extensive wiring, which limit their practical deployment. In contrast, Static Array Reconfiguration (SAR) is a one time, hardware simple, and cost effective solution requiring no real time control or computation. Although SAR cannot adapt to rapidly changing shading, its simplicity has motivated extensive research into improving PV performance under partial shading.
}

\textcolor{black}{
SuDoKu-based \cite{rani2013enhanced, sagar2020ku} and optimal SuDoKu \cite{krishna2019optimal} techniques are sensorless, mathematically driven SAR approaches that achieve effective shade dispersion, but their complex interconnections result in high wiring losses. Zigzag reconfiguration \cite{vijayalekshmy2016novel} provides greater flexibility, while fixed reconfiguration \cite{horoufiany2019new} and column wise PV module reallocation (CIA) \cite{pillai2018simple} focus on reducing wiring losses, though these methods suffer from limited scalability. A prime number based arrangement \cite{rezazadeh2021novel} enhances power output but relies on predictive data, limiting practical applicability.
}

\textcolor{black}{
Puzzle inspired methods such as Dominance Square (DS) \cite{dhanalakshmi2018dominance} and Magic Square (MS) \cite{yadav2017performance} offer either moderate shade dispersion or complex configurations \cite{sharma2023review}. Other approaches, including Lo Shu-based \cite{venkateswari2020power}, backtracking based \cite{tatabhatla2020performance}, skyscraper algorithms \cite{nihanth2019enhanced}, and Futoshiki reconfiguration \cite{sahu2015maximizing}, improve power output but introduce high computational complexity, excessive cabling losses, or sensitivity to irradiance variations.
}
\textcolor{black}{
Shape and pattern based SAR techniques such as L-shaped \cite{srinivasan2021shape}, Knight’s move based \cite{rezazadeh2022photovoltaic}, twisted two-step \cite{krishnan2022twisted}, and generalized repositioning \cite{singh2023optimal}, spiral pattern \cite{cherukuri2021power} have also been proposed. However, these methods do not consistently guarantee optimal solutions and fail to address diverse time varying and seasonal shading patterns, thereby limiting maximum power extraction in practical SPV systems.
}\\
\textcolor{black}{
This study introduces a novel SAR technique that explicitly accounts for the actual shading pattern experienced by the PV array. The proposed method accounts for shading effects from nearby fixed obstructions that vary with the position of the sun over different times of the day and across seasons, enabling consistent extraction of optimal power from the PV array under partial shading conditions. The performance of the proposed approach is compared with several existing reconfiguration techniques under different partial shading scenarios. Furthermore, its practical feasibility and real world applicability are validated through the development and experimental testing of a small scale laboratory prototype.
}

\textcolor{black}{
The rest of this article is organized as follows; the methodology of the proposed integer linear programming is discussed in Section \ref{Methodology}. Section \ref{lab:MILP_SAR_experimental_outcome} and Section \ref{lab:Hardware_exp} present the simulation and hardware experimental outcomes respectively. Finally, Section \ref{sec:Conclussion} concludes the study.
}
\begin{figure*}[!ht]
    \centering
    \begin{subfigure}[a]{0.3\textwidth}
		\includegraphics[width=\textwidth]{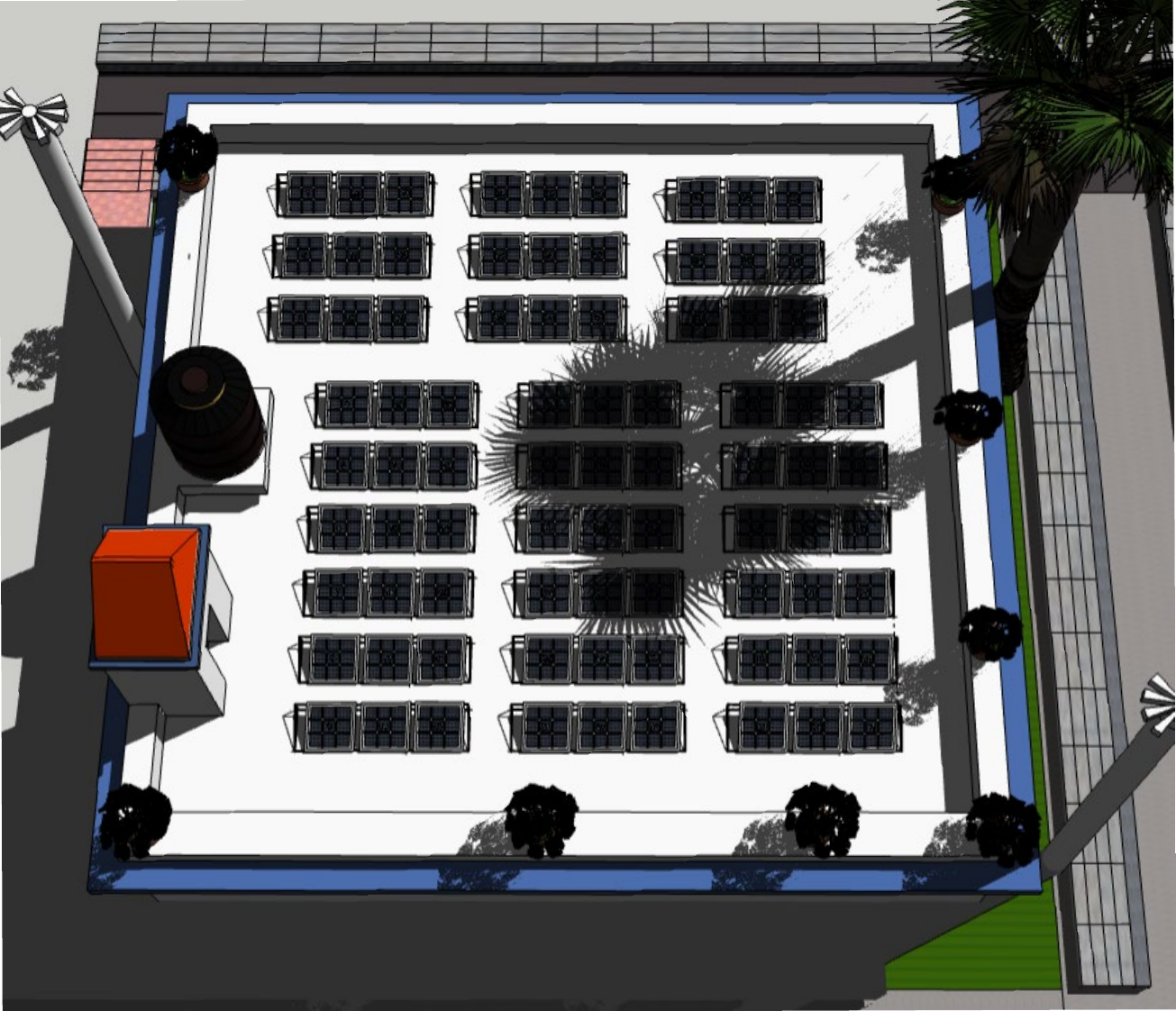}
		\caption{08:30}
		\label{fig:PF_29_4_7_30} \end{subfigure}
    \begin{subfigure}[c]{0.3\textwidth}
		\includegraphics[width=\textwidth]{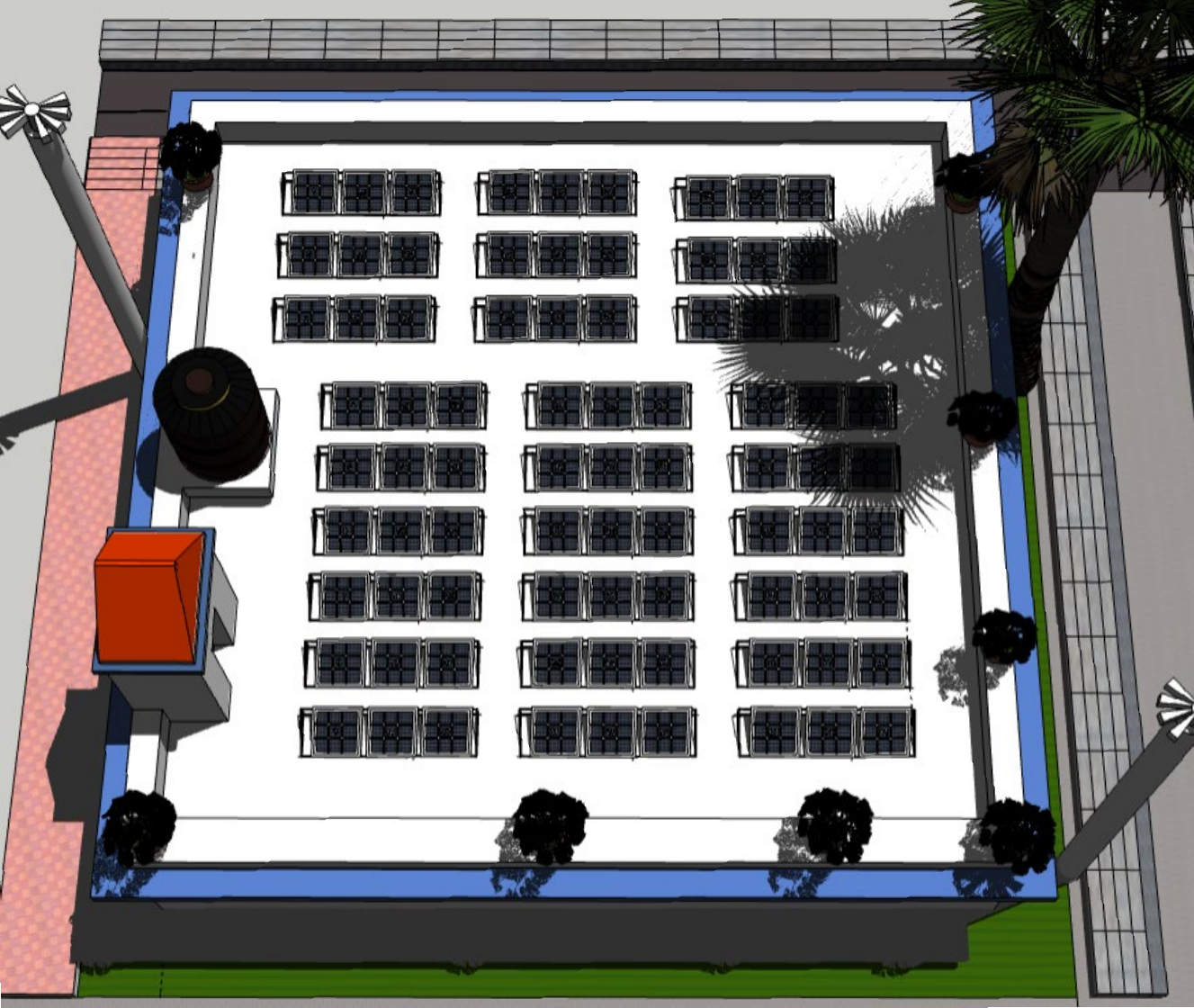}
		\caption{10:00}
		\label{fig:PF_29_4_14_32}	\end{subfigure}
    \begin{subfigure}[c]{0.3\textwidth}
		\includegraphics[width=\textwidth]{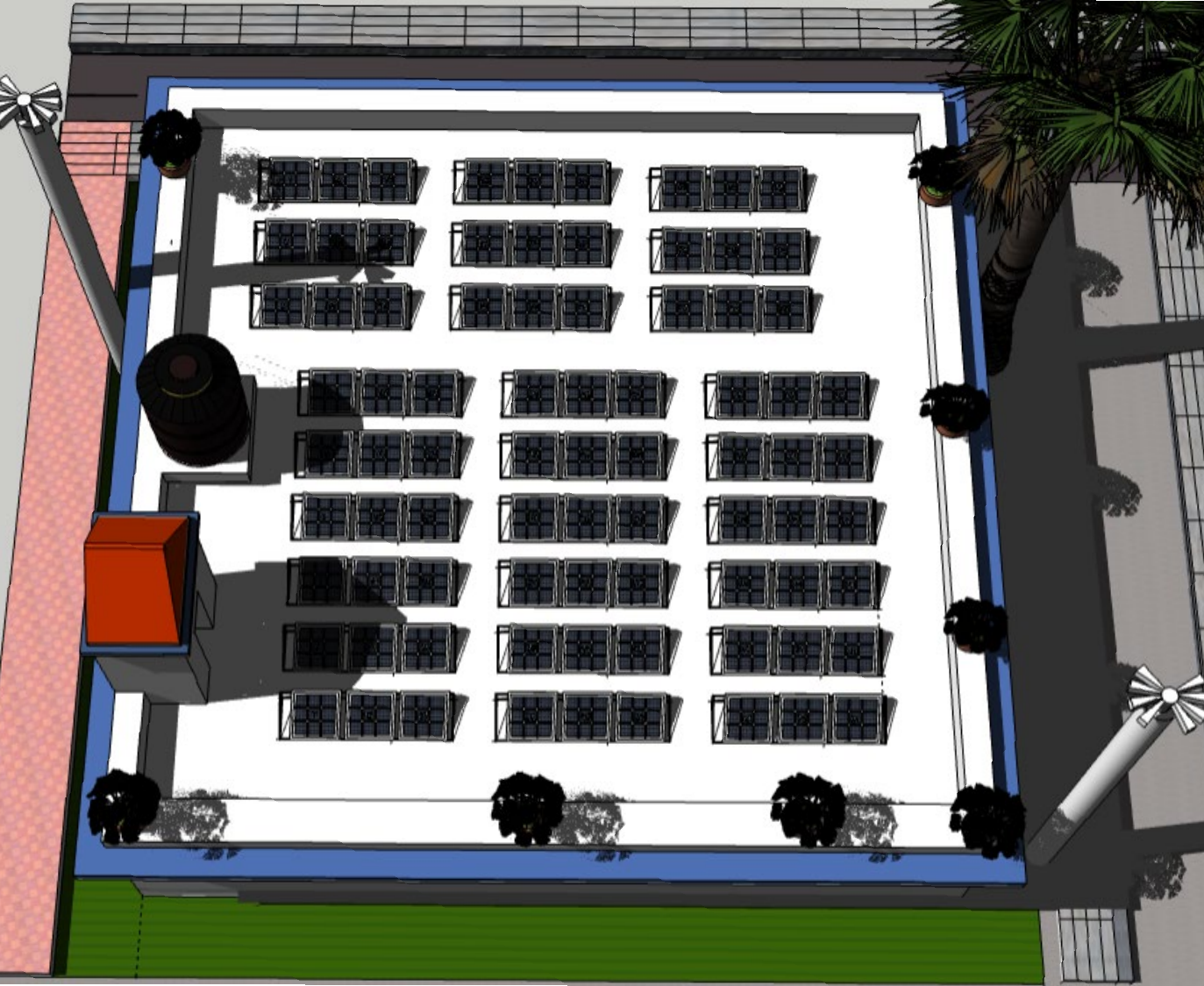}
		\caption{14:30}
		\label{fig:PF_29_4_15_30}	\end{subfigure}
    \caption{\textcolor{black}{Shading of different module at various time on the $166th$ day of year}}
    \label{fig:Problem_formulation_29_04} 
\end{figure*}

\begin{figure*}[!ht]
    \centering
    \begin{subfigure}[a]{0.3\textwidth}
		\includegraphics[width=\textwidth]{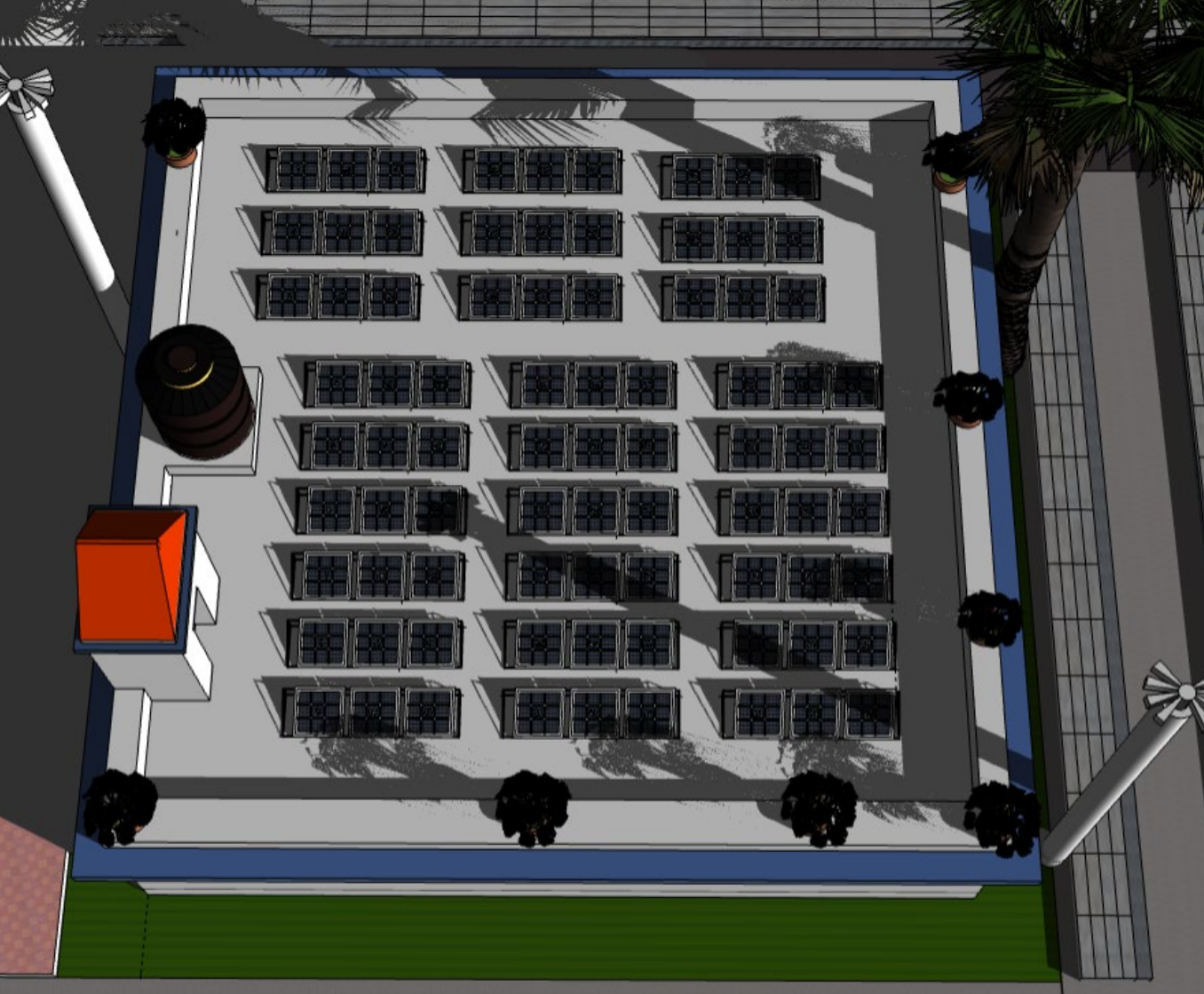}
		\caption{08:30}
		\label{fig:PF_28_10_7_30} \end{subfigure}
    \begin{subfigure}[c]{0.3\textwidth}
		\includegraphics[width=\textwidth]{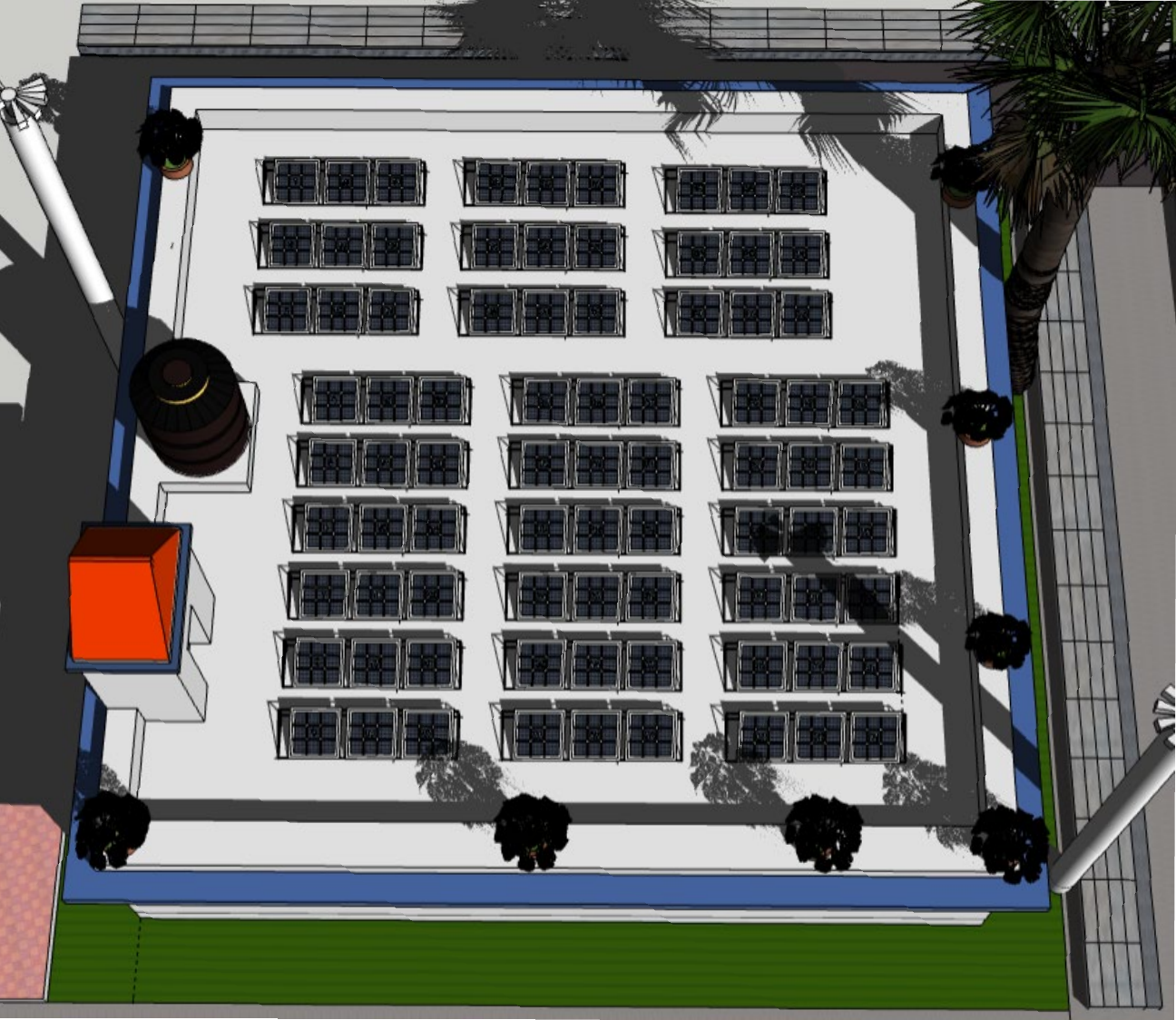}
		\caption{10:00}
		\label{fig:PF_28_10_14_32}	\end{subfigure}
    \begin{subfigure}[c]{0.3\textwidth}
		\includegraphics[width=\textwidth]{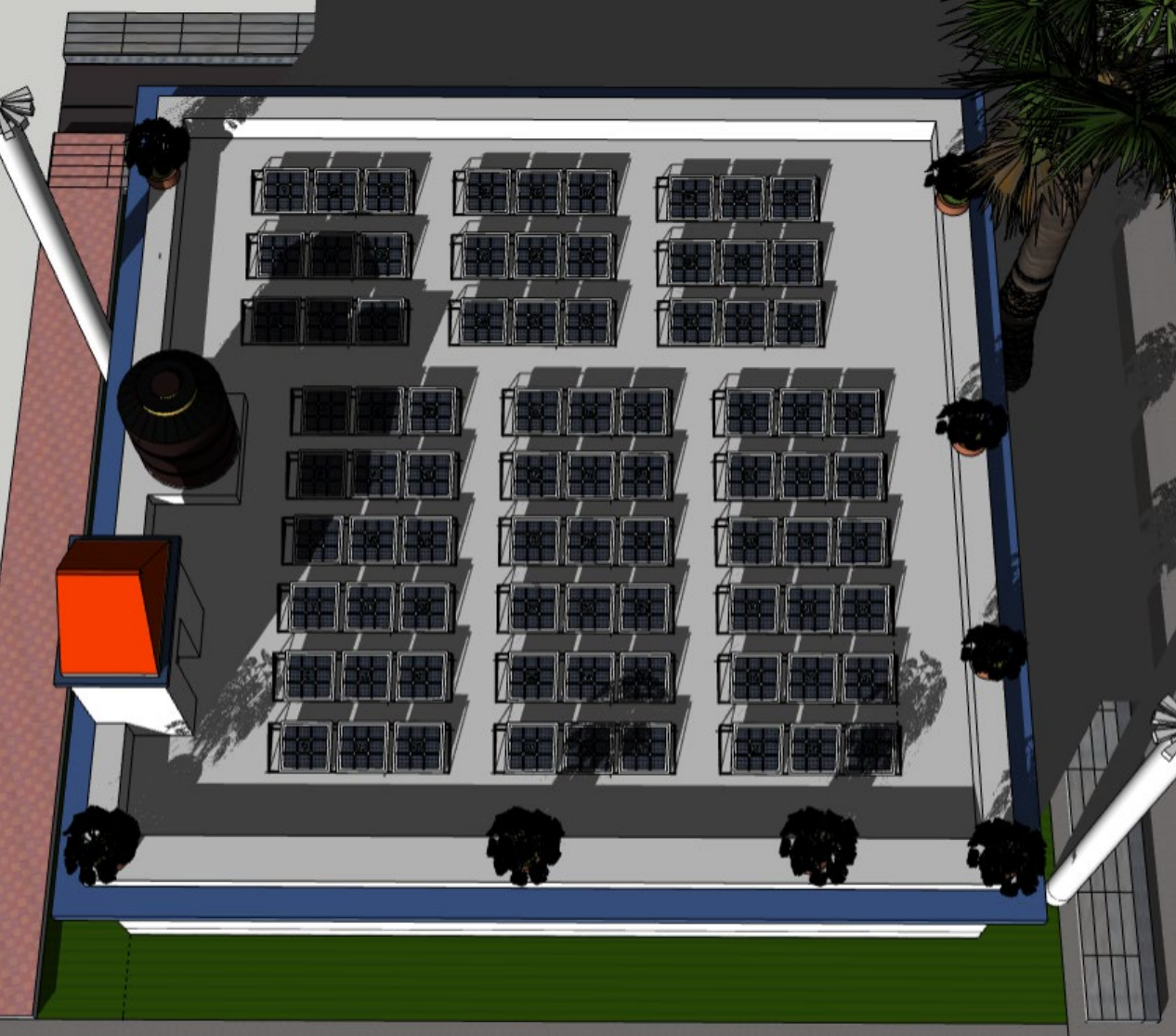}
		\caption{14:30}
		\label{fig:PF_28_10_15_30}	\end{subfigure}
    \caption{\textcolor{black}{Shading of different module at various time on the $349th$ day of year}}
    \label{fig:Problem_formulation_28_10} \vspace{-3ex}
\end{figure*}

\section{Methodology} \label{Methodology}
\subsection{Problem Description} \label{sec:Problem Formulation} 

\textcolor{black}{
Solar PV plants deployed in urban environments are frequently subjected to partial shading due to surrounding high rise buildings, trees, and tall utility or lighting poles. These obstructions cause non uniform irradiance across the SPV array, resulting in spatially and temporally varying shading patterns. Such conditions lead to current mismatch among PV modules, increased power losses, and thereby reducing overall energy extraction. Under these operating conditions, appropriate PV module reconfiguration becomes essential to ensure stable and optimal power generation.
}

\textcolor{black}{
A single static reconfiguration of the PV array at the time of installation is inadequate for sustaining maximum power generation throughout the year because the irradiance distribution on the SPV modules changes dynamically with seasonal changes. This non-uniformity arises from variations in solar azimuth and altitude, which cause the relative movement of shadows cast by nearby obstructions across the PV array at different times of the day and across seasons. This phenomenon is illustrated in \figurename{~\ref{fig:Problem_formulation_29_04}} and \figurename{~\ref{fig:Problem_formulation_28_10}}, where a $9\times 9$ SPV array is installed on the rooftop of a single-story building. There is a palm tree on the east side of the plant, one electric pole is situated in the southeast and another one on the west side, the attic is positioned in the southwest direction, and a water tank is on the west. The building is decorated with small plants in fixed pots.   
}

\textcolor{black}{
The entire simulation environment was developed using Google SketchUp to analyze the daily and seasonal shading behavior caused by these obstructions.
\figurename{~\ref{fig:Problem_formulation_29_04}} and \figurename{~\ref{fig:Problem_formulation_28_10}} illustrate the shading patterns observed at different times on the 166th and 349th days of the year, respectively, as these days are proximate to the solstices. At $08:30$, \figurename{~\ref{fig:PF_29_4_7_30}} and \figurename{~\ref{fig:PF_28_10_7_30}} show that $22$ and $17$ modules are shaded, respectively. At $10:00$, analysis of \figurename{~\ref{fig:PF_29_4_14_32}} and \figurename{~\ref{fig:PF_28_10_14_32}} indicates shading of $8$ and $6$ modules, while at $14:30$, \figurename{~\ref{fig:PF_29_4_15_30}} and \figurename{~\ref{fig:PF_28_10_15_30}} reveal shading of $10$ and $17$ modules on the $166$th and $349$th days, respectively.
Owing to the wide variations in shading patterns, it is challenging to identify a single configuration that can extract optimal power from the PV array under all partial shading conditions occurring throughout the day and across seasons.
}
\vspace{-2ex}
\begin{figure}[t]
    \centering
	\includegraphics[width=0.4\textwidth]{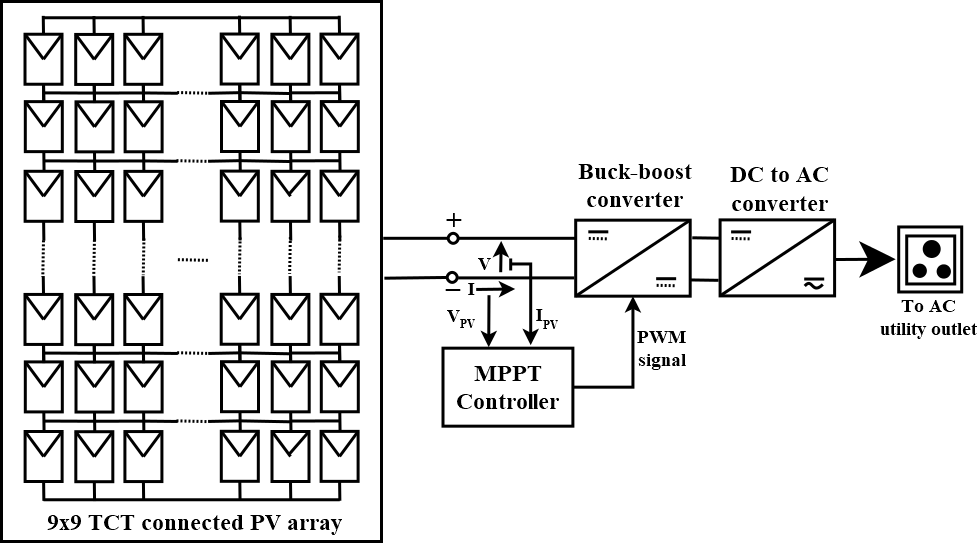}
	\caption{\textcolor{black}{Schematic diagram of TCT connected PV array system}}
    \label{fig:schematic_diagram} \vspace{-3ex}
\end{figure}
\subsection{Integer linear programming algorithm to extract maximum row current}\label{sec:MILP_row_wise}

\textcolor{black}{
As discussed in the previous section, PV array shading varies throughout the day and across seasons due to changes in the sun’s position. As a result, different PV modules are shaded under different shading scenarios. The proposed static reconfiguration technique electrically relocates each module in a TCT-configured PV array only once, with the objective of minimizing the number of shaded modules in any row for all shading patterns occurring throughout the year. This ensures maximization of the array output current and, consequently, the overall power output under all considered shading conditions. Short-duration shading due to clouds is neglected, while persistent shading caused by fixed obstacles such as trees, nearby buildings, street lights and water tanks are considered.}\\

\noindent{\textbf{\textit{Mathematical Formulation using ILP:}}}

\textcolor{black}{Annual shading patterns on the PV array are monitored, with each pattern shading a specific set of modules. The objective is to identify a unique, static PV configuration that distributes these shaded modules as evenly as possible across all rows for every recorded pattern. To achieve this, the challenge is formulated as a robust job scheduling problem designed to handle multiple shading scenarios with a single arrangement.
In this model, assigning a physical position to a PV module is equivalent to assigning a job to a machine (a row). This assignment ($X_{j,k}$) remains static across all shading patterns. However, the \textit{load} on each machine fluctuates based on which modules become active (shaded) during a specific pattern.
For every shading pattern $o$, we determine the completion time $T_o$, defined as the maximum number of shaded modules in any single row. The ultimate goal is to find the optimal arrangement ($X_{j,k}$) that minimizes the sum of completion times across all defined patterns. This ensures the array is robustly optimized to handle every shading scenario effectively.}


\begin{itemize}
    \item \textbf{Single Schedule:} The decision variable $X_{j,k}$ determines the fixed physical position of module $j$ in row $k$. This arrangement remains constant across all shading patterns.
    \item \textbf{Multiple Objectives:} Each shading pattern represents a distinct sub-objective. For any given pattern $o$, only the specific subset of shaded modules, denoted as $Jobs_o$, contributes to the load on a given machine (row).
    \item \textbf{Goal:} Minimize the total cumulative completion time across all sub-objectives.
\end{itemize}

\subsection*{1. Sets}
\begin{itemize}
    \item $\mathcal{J} = \{1, 2, \dots, P\}$: Set of all shaded PV modules (jobs) in the array.
    \item $\mathcal{M} = \{1, 2, \dots, m\}$: Set of all rows (machines).
    \item $\mathcal{O} = \{1, 2, \dots, q\}$: Set of all shading patterns (objectives).
\end{itemize}
\vspace{-2ex}
\subsection*{2. Parameters}
\begin{itemize}
    \item $P$: Total number of jobs (shaded modules).
    \item $m$: Total number of machines (rows).
    \item $n$: Maximum capacity of a row (number of columns in the array).
    \item $Jobs_o \subset \mathcal{J}$: The subset of jobs that are shaded (active) during shading pattern $o \in \mathcal{O}$.
\end{itemize}
\subsection*{3. Decision Variables}
\begin{itemize}
    \item $X_{j,k} \in \{0, 1\}$: Binary variable. Equal to 1 if job $j$ is assigned to machine $k$. This assignment is \textbf{fixed} for all objectives.
    \item $N_{o,k}^{\text{obj}} \in \mathbb{Z}_{\ge 0}$: Integer variable. The number of shaded modules from pattern $o$ that end up on machine $k$ based on the fixed assignment $X_{j,k}$.
    \item $T_o \in \mathbb{Z}_{\ge 0}$: Integer variable. The completion time (maximum row load) for shading pattern $o$.
\end{itemize}
\subsection*{4. Objective Function}
The overall objective is to minimize the sum of completion times for all sub-objectives. This drives the solver to find a single configuration $X_{j,k}$ that results in a low completion time for every individual shading pattern.
\begin{align*}
    \text{Minimize } & Z = \sum_{o \in \mathcal{O}} T_o
\end{align*}
\vspace{-4ex}
\subsection*{5. Constraints}
The problem is subject to the following constraints:
\begin{align}
    \sum_{k \in \mathcal{M}} X_{j,k} &= 1 && \forall j \in \mathcal{J} \label{eq:static_assignment} \\
    \sum_{j \in \mathcal{J}} X_{j,k} &\le n && \forall k \in \mathcal{M} \label{eq:physical_capacity} \\
    N_{o,k}^{\text{obj}} &= \sum_{j \in Jobs_o} X_{j,k} && \forall o \in \mathcal{O}, \forall k \in \mathcal{M} \label{eq:pattern_load} \\
    T_o &\ge N_{o,k}^{\text{obj}} && \forall o \in \mathcal{O}, \forall k \in \mathcal{M} \label{eq:calc_To} \\
    X_{j,k} &\in \{0, 1\} && \forall j \in \mathcal{J}, \forall k \in \mathcal{M} \\
    N_{o,k}^{\text{obj}}, T_o &\in \mathbb{Z}_{\ge 0} && \forall o \in \mathcal{O}, \forall k \in \mathcal{M}
\end{align}

\begin{itemize}
    \item Constraint \eqref{eq:static_assignment} ensures every module $j$ is assigned to exactly one row $k$. This defines the single static schedule.
    \item Constraint \eqref{eq:physical_capacity} ensures the physical capacity of the rows is not exceeded (total modules per row $\le n$).
    \item Constraint \eqref{eq:pattern_load} calculates the specific load for pattern $o$ on row $k$. It sums $X_{j,k}$ \textbf{only} for the jobs involved in that specific pattern ($j \in Jobs_o$).
    \item Constraint \eqref{eq:calc_To} determines the completion time $T_o$ for pattern $o$. Since we minimize $\sum T_o$, this inequality forces $T_o$ to equal the maximum load ($\max_k N_{o,k}^{\text{obj}}$) for that pattern.
    \item Constraint (5) defines $X_{j,k}$ as a binary decision variable. It restricts the values to either 0 or 1, ensuring that a module $j$ is either assigned to row $k$ ($X=1$) or not assigned to it ($X=0$).
    \item Constraint (6) enforces integrality and non-negativity for the auxiliary variables. 
\end{itemize}

\section{Simulation Experiment} \label{lab:MILP_SAR_experimental_outcome}
\textcolor{black}{
This section discusses the computation of a single optimal configuration for the $9\times 9$ rooftop PV array illustrated in \figurename{~\ref{fig:Problem_formulation_29_04}} and \figurename{~\ref{fig:Problem_formulation_28_10}}. As described in Section~\ref{sec:Problem Formulation}, the physical installation of the $9\times 9$ PV array was modeled using Google SketchUp to capture the shading effects of nearby obstructions throughout the day and across different seasons. Based on this model, a total of $54$ distinct shading patterns were identified, each affecting a different subset of PV modules.
}

\begin{figure}[t]
    \centering
	\includegraphics[width=0.4\textwidth]{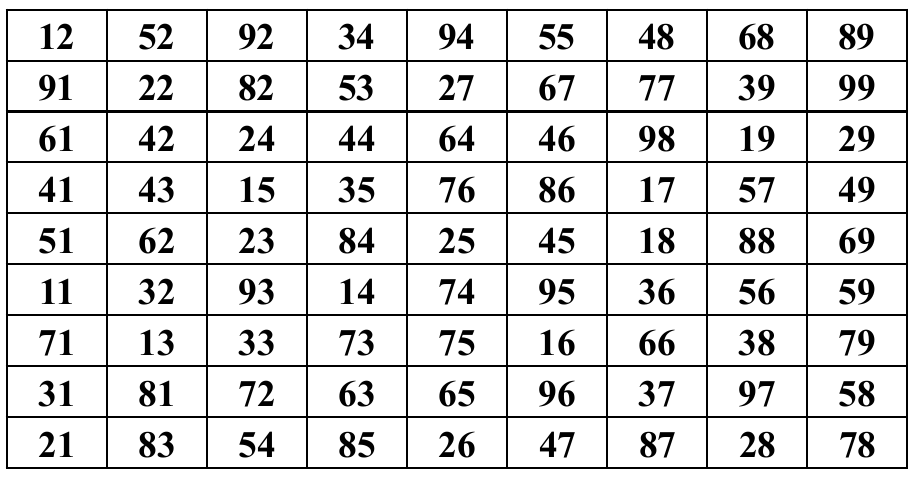}
	\caption{\textcolor{black}{Reconfigured ILP-based PV array}}
    \label{fig:ERI_Proposed_MILP} \vspace{-3ex}
\end{figure}

\textcolor{black}{
The proposed ILP based reconfiguration method, detailed in Section~\ref{sec:MILP_row_wise}, is then applied to these shading patterns to determine a single static configuration that mitigates the impact of partial shading. The resulting reconfigured PV array, indicating the electrical positions of the PV modules, is shown in \figurename{~\ref{fig:ERI_Proposed_MILP}}.
}


\textcolor{black}{
The electrical simulation of the proposed ILP-based SAR technique is carried out in MATLAB for the $9\times 9$ SPV array. The schematic diagram of the simulation is shown in \figurename{~\ref{fig:schematic_diagram}}. The PV module specifications for the simulations and practical experiments are presented in \tablename{~\ref{tab: Module specification}}. In the simulation, the performance of the proposed ILP-based SAR under partial shading was also compared with the conventional TCT, SuDoKu \cite{sagar2020ku}, Optimal SuDoKu \cite{krishna2019optimal}, TTSA \cite{krishnan2022twisted}, GMRA \cite{singh2023optimal}, and SPA \cite{cherukuri2021power} reconfiguration methods. Two different shading cases in two different seasons, $166th$ and $349th$ days of the year are considered for carrying out the simulation. The fill factor (FF), percentage mismatch loss (PML), and execution ratio (ER) are the SPV performance indices used to compare results.
}


\begin{table}[ht]
    \centering
     \caption{Solar PV module specification}
    \label{tab: Module specification}
    \begin{tabular}{|p{3.7cm}|p{1.8cm}| p{2.0cm}|}
        \hline
        \textbf{Parameters} & \textbf{For simulation} & \textbf{for experiment}\\
        \hline
        Manufacturer Name & Apollo Solar Energy ASEC-145G6M49 & Solar Solutions India\\
        \hline
        Maximum Power & $144.97~W$ &  $10~W$ \\
        \hline
        Open Circuit Voltage ($V_{OC}$) & $22.32~V$ & $21.0~V$\\
        \hline
        Short Circuit Current ($I_{SC}$) & $8.66~A$ & $0.61~A$\\
        \hline
        Voltage at maximum power point ($V_{MPP}$) & $17.81~V$ & $17.3~V$\\
        \hline
        Current at maximum power point ($I_{MPP}$) & $8.14~A$ & $0.57~A$\\
        \hline
    \end{tabular}
   \end{table}
\vspace{-2ex}

\vspace{-2ex}
\subsection{Case-I: $166th$ day of year} \label{lab:case_1}
\begin{figure}[htbp]
    \centering
     \begin{subfigure}[a]{0.4\textwidth}
		\includegraphics[width=\textwidth]{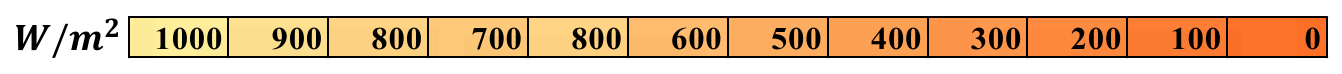}
		\label{fig:job_irr_2} \end{subfigure}\\ \vspace{-2ex}
    \begin{subfigure}[a]{0.36\textwidth}
		\includegraphics[width=\textwidth]{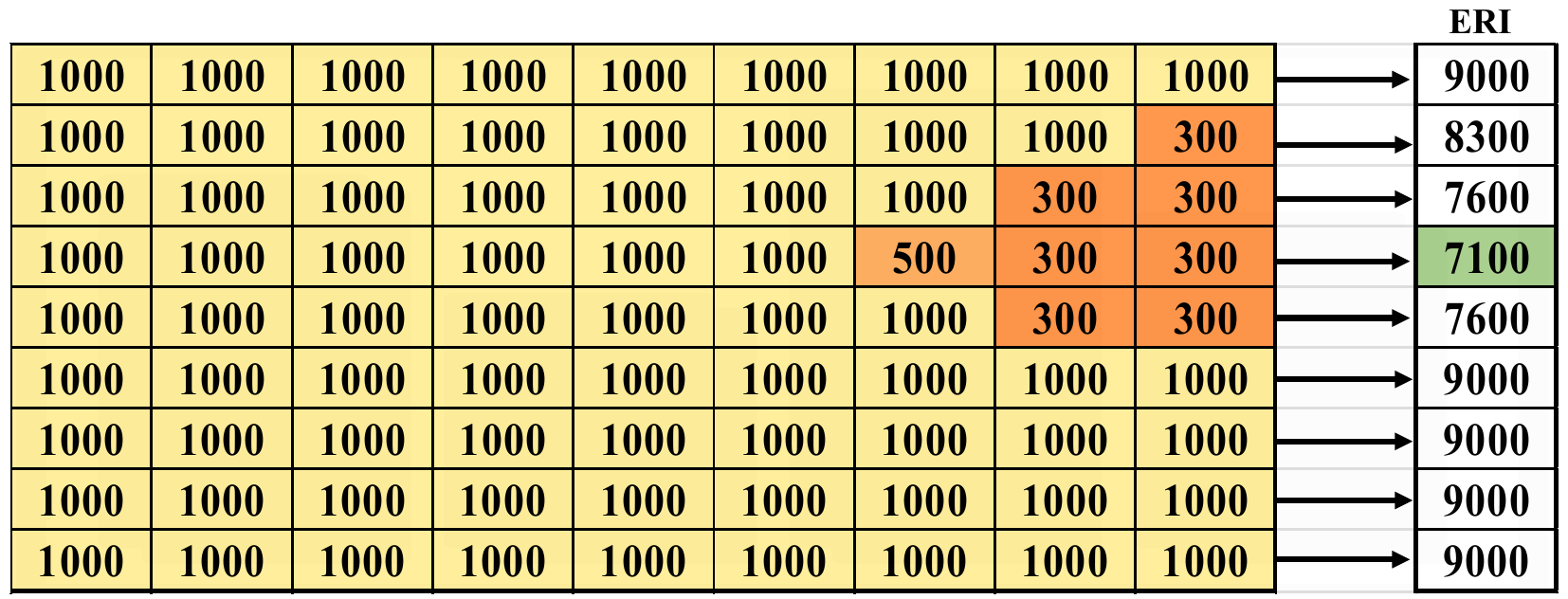}
		\caption{TCT}
		\label{fig:job_48_TCT} \end{subfigure}
    \begin{subfigure}[c]{0.36\textwidth}
		\includegraphics[width=\textwidth]{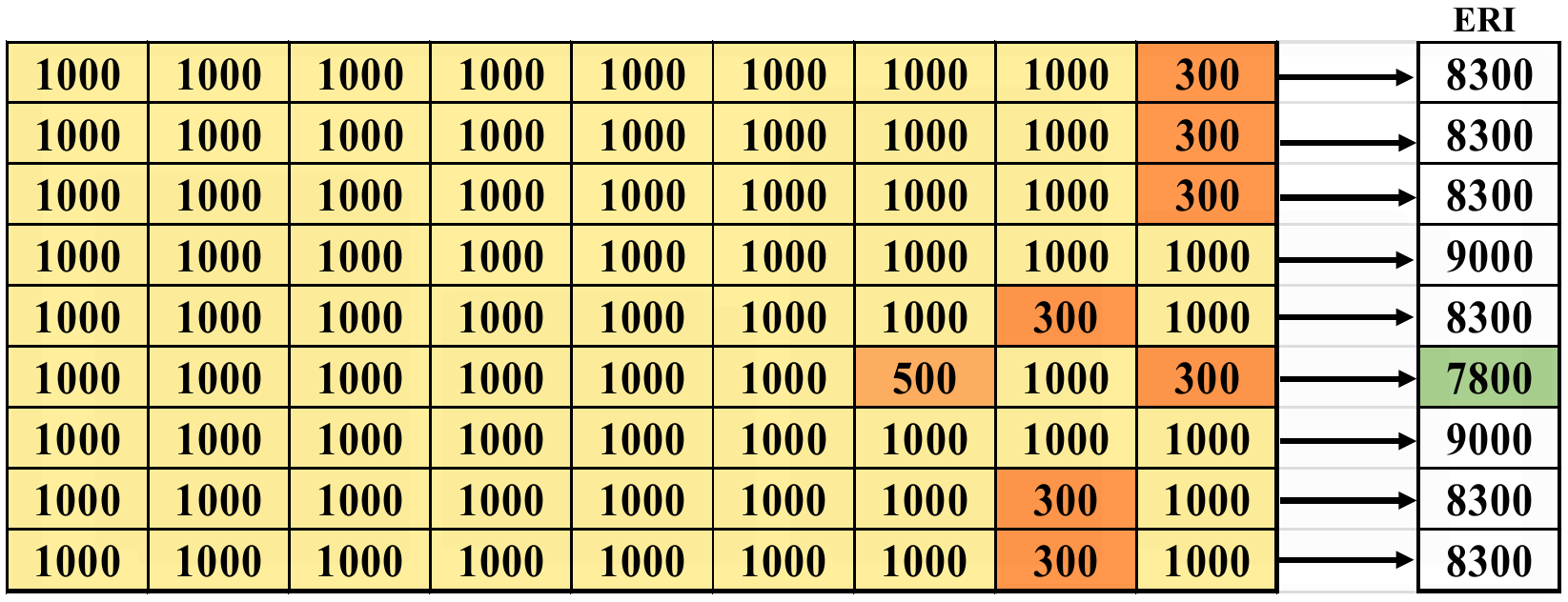}
		\caption{SuDoKu}
		\label{fig:fig:job_48_SUDOKU}	\end{subfigure}
    \begin{subfigure}[c]{0.36\textwidth}
		\includegraphics[width=\textwidth]{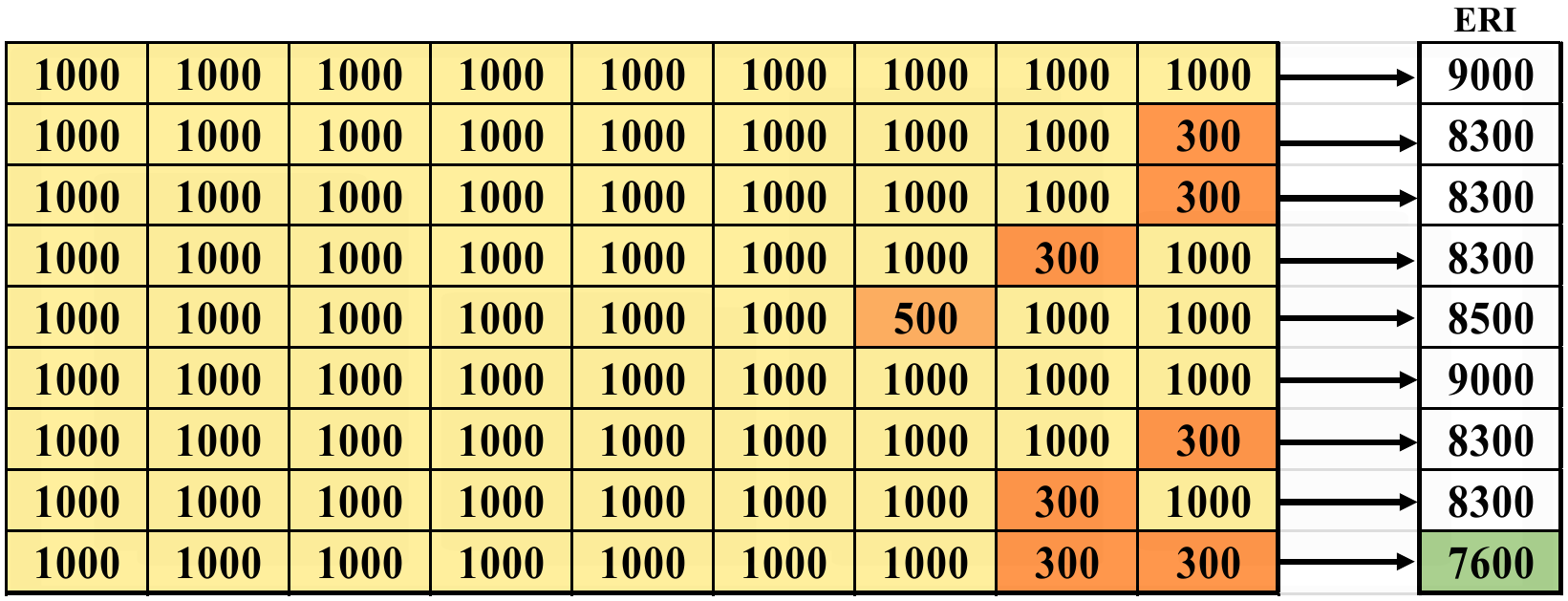}
		\caption{Optimal SuDoKu}
		\label{fig:fig:job_48_OPT_SUDOKU}	\end{subfigure}
    \begin{subfigure}[a]{0.36\textwidth}
		\includegraphics[width=\textwidth]{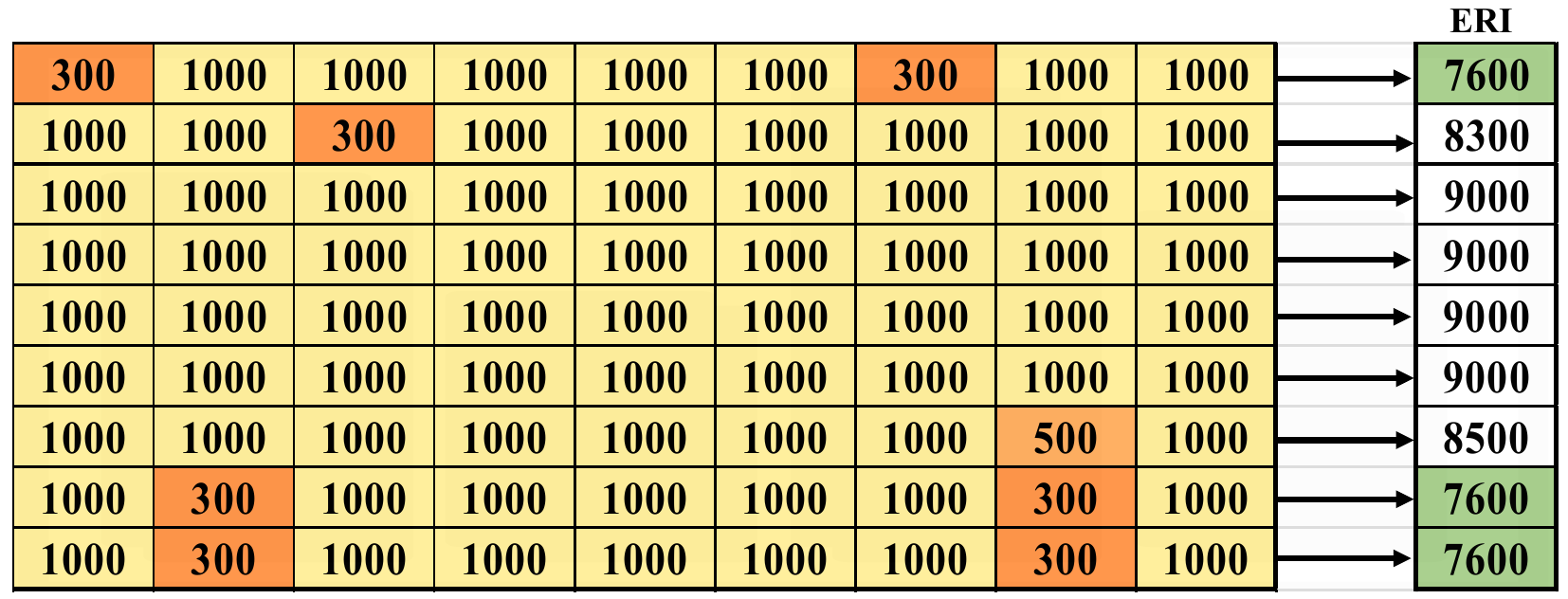}
		\caption{TTSA}
		\label{fig:job_48_TTSA} \end{subfigure}
    \begin{subfigure}[c]{0.36\textwidth}
		\includegraphics[width=\textwidth]{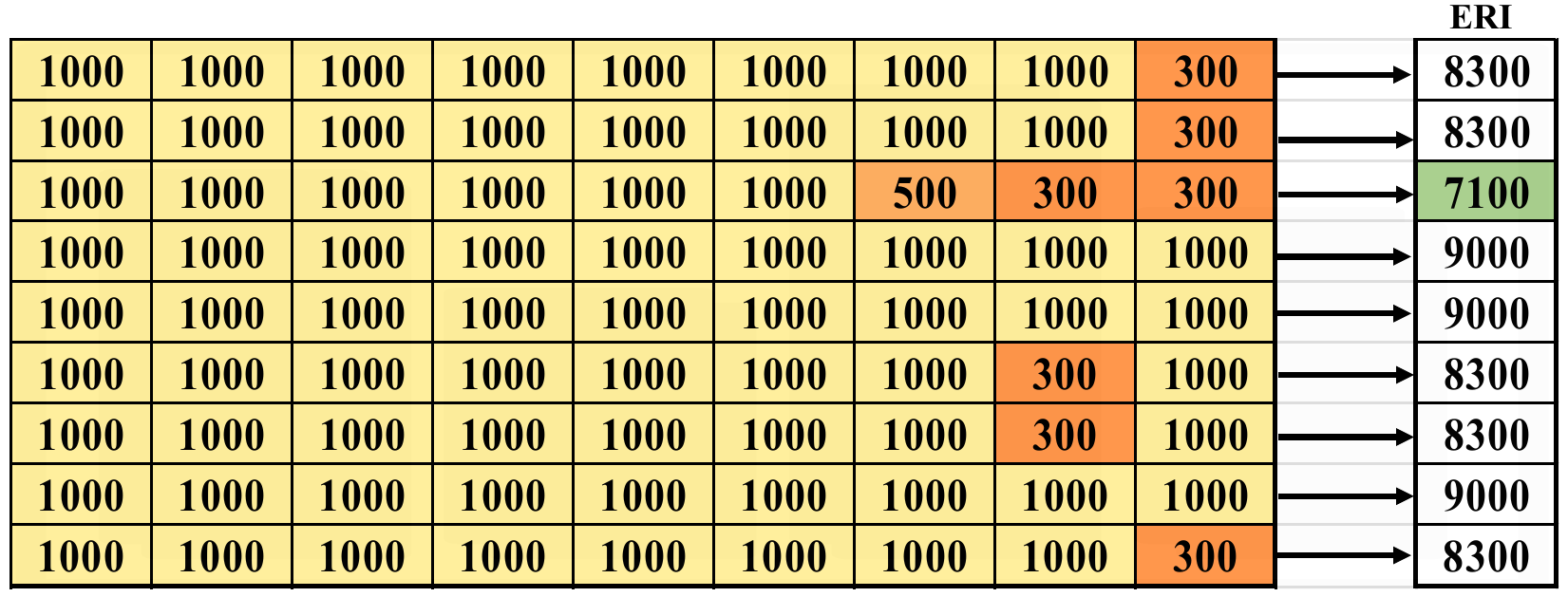}
		\caption{GMRA}
		\label{fig:fig:job_48_GMRA}	\end{subfigure}
    \begin{subfigure}[a]{0.36\textwidth}
		\includegraphics[width=\textwidth]{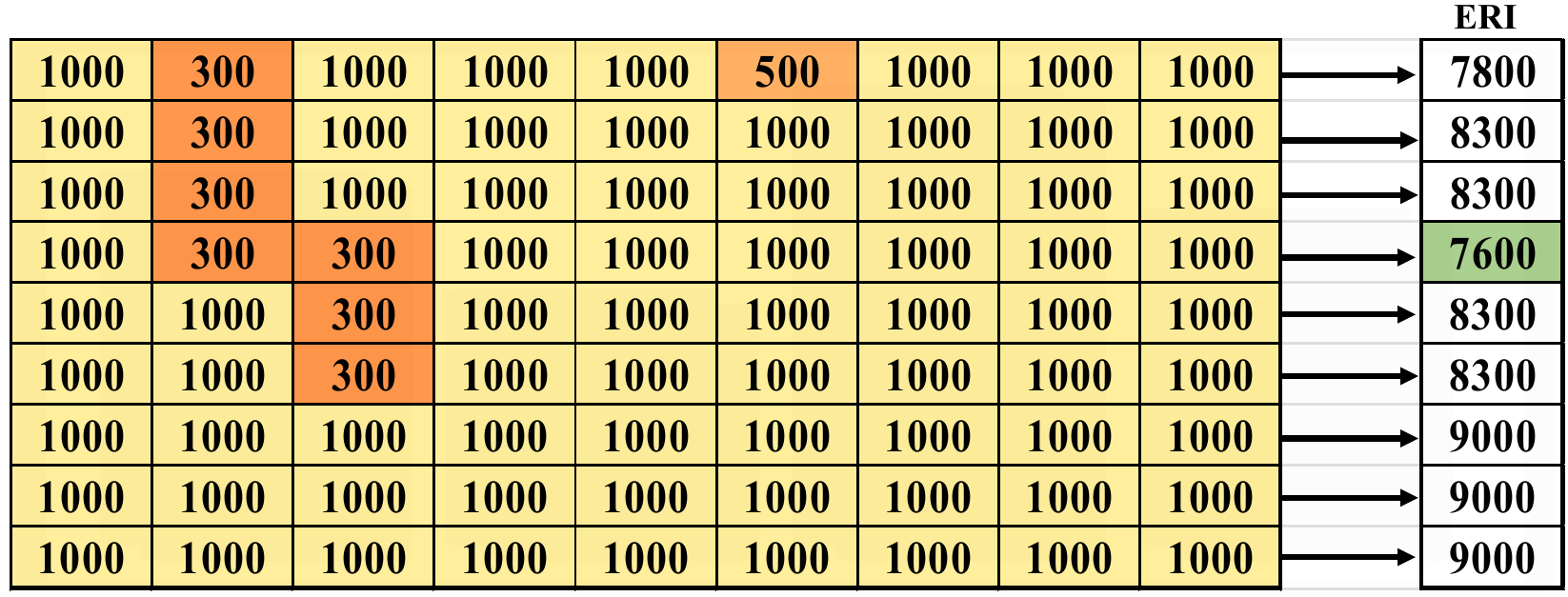}
		\caption{SPA}
		\label{fig:job_48_SPA} \end{subfigure}
    \begin{subfigure}[a]{0.36\textwidth}
		\includegraphics[width=\textwidth]{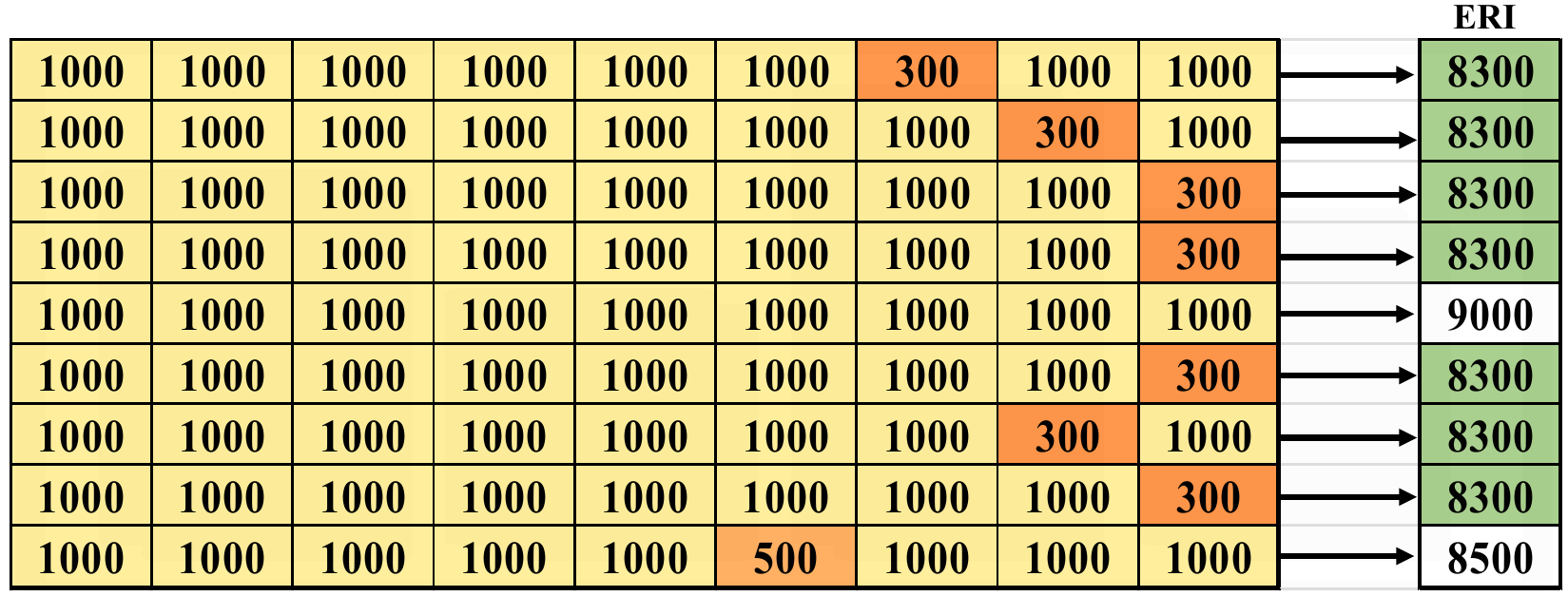}
		\caption{ILP}
		\label{fig:job_48_MILP} \end{subfigure}
    \caption{\textcolor{black}{Case-I: The shade dispersion of a $9\times 9$ SPV array at $10:00$ of the $166th$ day of year with (a) TCT, (b) SuDoKu \cite{sagar2020ku}, (c) Optimal SuDoKu \cite{krishna2019optimal}, (d) TTSA \cite{krishnan2022twisted}, (e) GMRA \cite{singh2023optimal}, (f) SPA \cite{cherukuri2021power}, and (g) Proposed ILP}}
    \label{fig:Shading_Patterns_29_04} 
\end{figure}

\textcolor{black}{
In this case, the shading pattern observed on the $166$st day of the year at $10:00$ was analyzed. At this instant, eight PV modules are affected by partial shading. The global irradiance was $1000~W/m^2$, while the shaded modules received non-uniform irradiance levels of $500~W/m^2$ and $300~W/m^2$. The shading distribution for the conventional TCT configuration, along with the corresponding effective row irradiance (ERI) in $W/m^2$, is illustrated in \figurename{~\ref{fig:job_48_TCT}}.
The shade dispersion patterns and corresponding ERI obtained using SuDoKu \cite{sagar2020ku}, Optimal SuDoKu \cite{krishna2019optimal}, TTSA \cite{krishnan2022twisted}, GMRA \cite{singh2023optimal}, SPA \cite{cherukuri2021power}, and the proposed ILP based SAR technique are shown in \figurename{~\ref{fig:fig:job_48_SUDOKU}}, \figurename{~\ref{fig:fig:job_48_OPT_SUDOKU}}, \figurename{~\ref{fig:job_48_TTSA}}, \figurename{~\ref{fig:fig:job_48_GMRA}}, \figurename{~\ref{fig:job_48_SPA}} and \figurename{~\ref{fig:job_48_MILP}}, respectively. From these figures, it is evident that the proposed ILP based SAR method achieves the highest minimum effective row irradiance (MERI) of $8300~W/m^2$.}

\begin{figure}[t]
    \centering
	\includegraphics[width=0.48\textwidth]{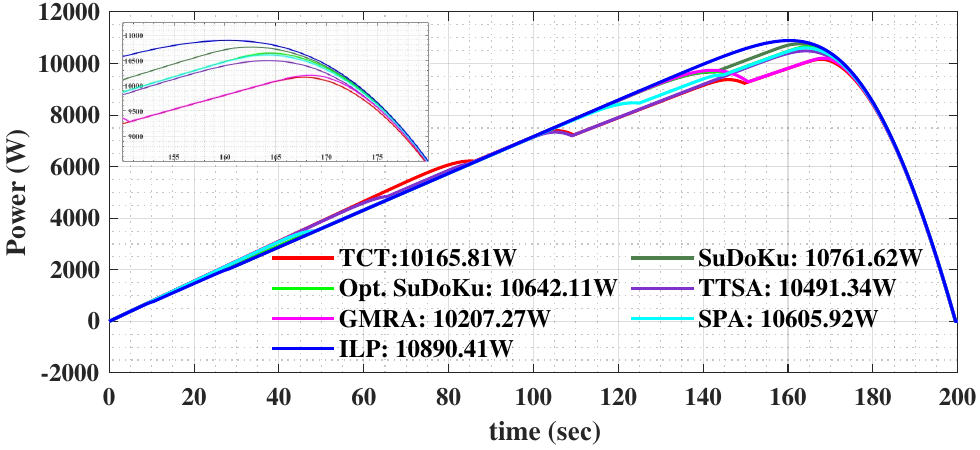}
	\caption{\textcolor{black}{Case-I: P-V characteristics for different reconfiguration strategies on the $166th$ day of year}}
    \label{fig:Case-I_P_V} \vspace{-3ex}
\end{figure}

\textcolor{black}{
The corresponding power-voltage (P-V) characteristics of the PV array under SuDoKu, Optimal SuDoKu, TTSA, GMRA, SPA, and the proposed ILP-based SAR configurations are presented in \figurename{~\ref{fig:Case-I_P_V}}. It can be observed that the proposed method enables the extraction of the highest output power of $10890.41~W$ under the considered partial shading condition.
}

\begin{figure}[ht]
    \centering
    \begin{subfigure}[a]{0.24\textwidth}
	\includegraphics[width=\textwidth]{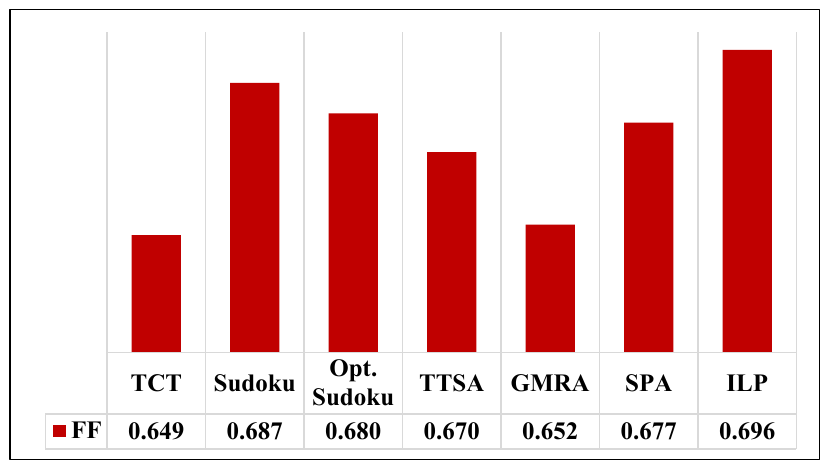}
	\caption{Fill Factor}
    \label{fig:Case-I_FF} \end{subfigure}
    \begin{subfigure}[a]{0.24\textwidth}
	\includegraphics[width=\textwidth]{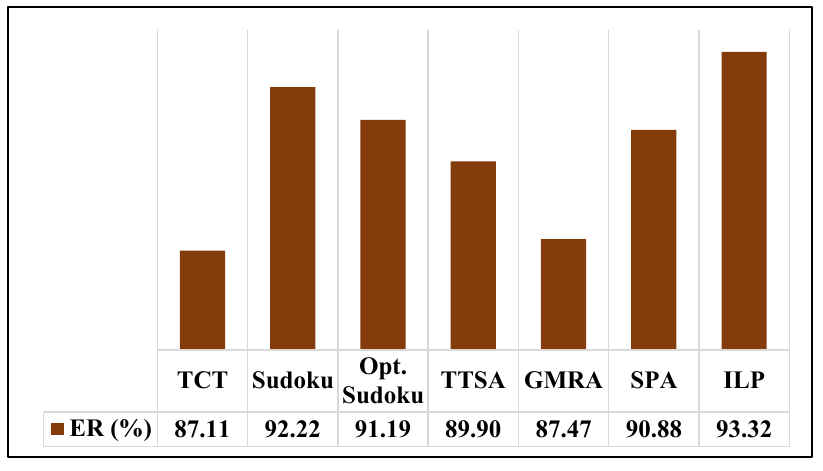}
	\caption{Execution Ratio}
    \label{fig:Case-I_ER} \end{subfigure}
    \begin{subfigure}[a]{0.24\textwidth}
    \includegraphics[width=\textwidth]{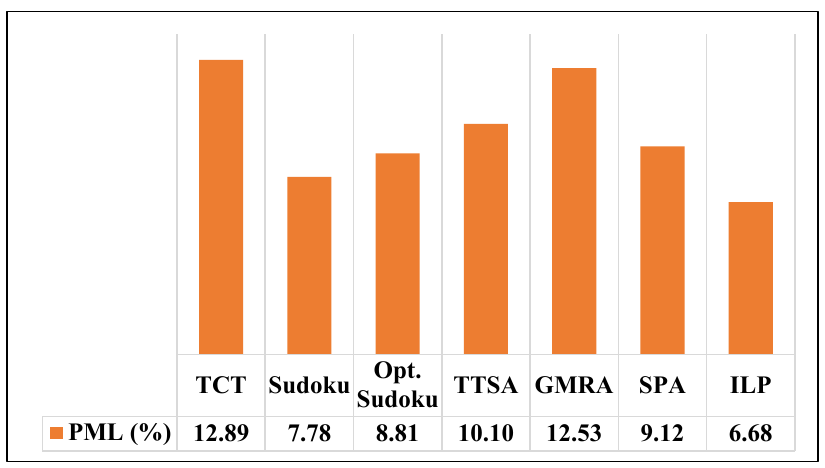}
	\caption{\% Mismatch Loss}
    \label{fig:Case-I_PML} \end{subfigure}
    \caption{Case-I: Performances indices of different array reconfiguration strategies on the $166th$ day of year} \label{fig:_PV_PI}
    \vspace{-3ex}
\end{figure}
\subsection{Case-II: $349th$ day of year} \label{lab:case_2} 
\begin{figure}[htbp]
    \centering
    \begin{subfigure}[a]{0.4\textwidth}
		\includegraphics[width=\textwidth]{./images/IRR_ROW.pdf}
		\label{fig:job_irr_1} \end{subfigure}\\ \vspace{-2ex}
    \begin{subfigure}[a]{0.36\textwidth}
		\includegraphics[width=\textwidth]{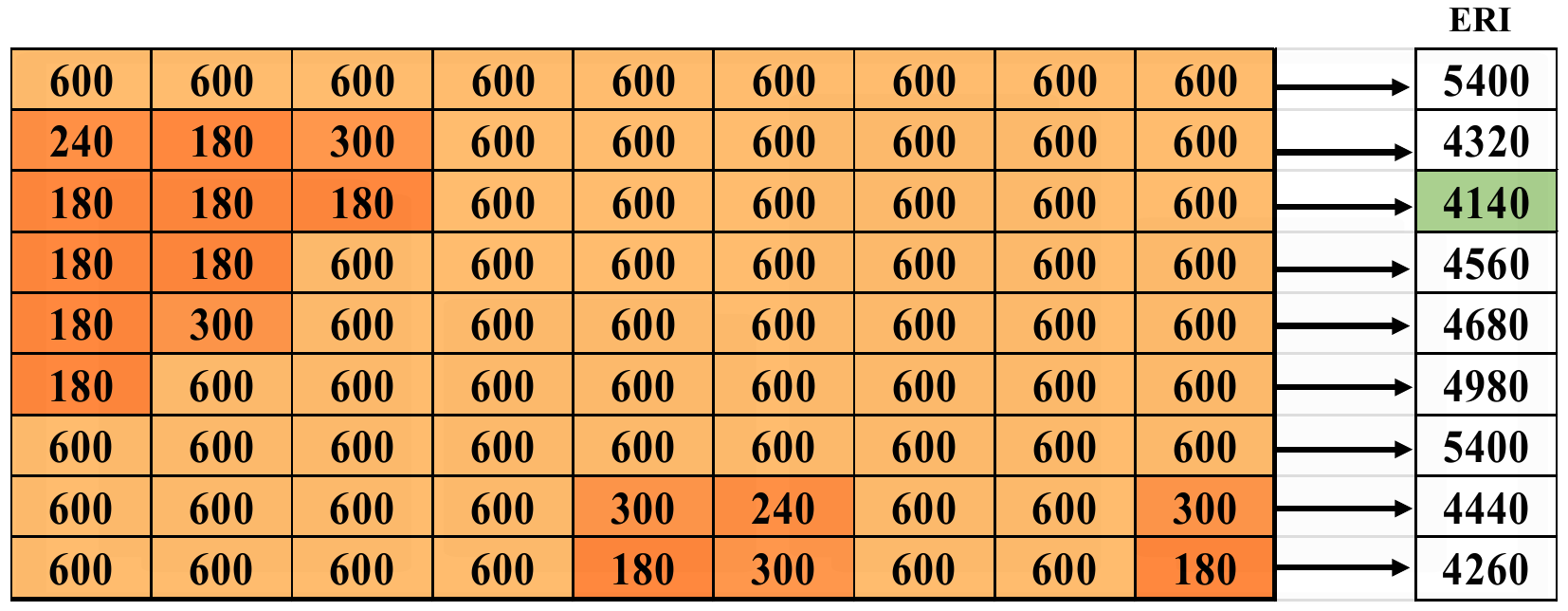}
		\caption{TCT}
		\label{fig:job_12_TCT} \end{subfigure}
    \begin{subfigure}[c]{0.36\textwidth}
		\includegraphics[width=\textwidth]{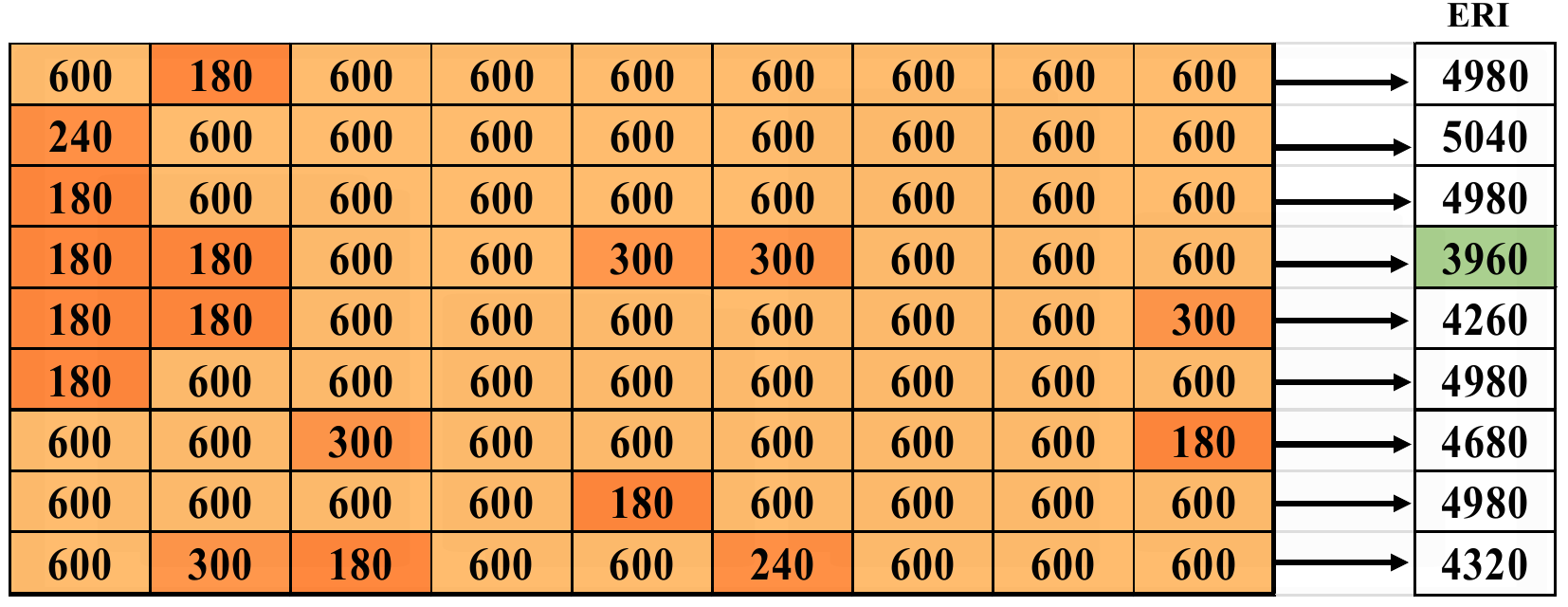}
		\caption{SuDoKu}
		\label{fig:fig:job_12_Sudoku}	\end{subfigure}
    \begin{subfigure}[c]{0.36\textwidth}
		\includegraphics[width=\textwidth]{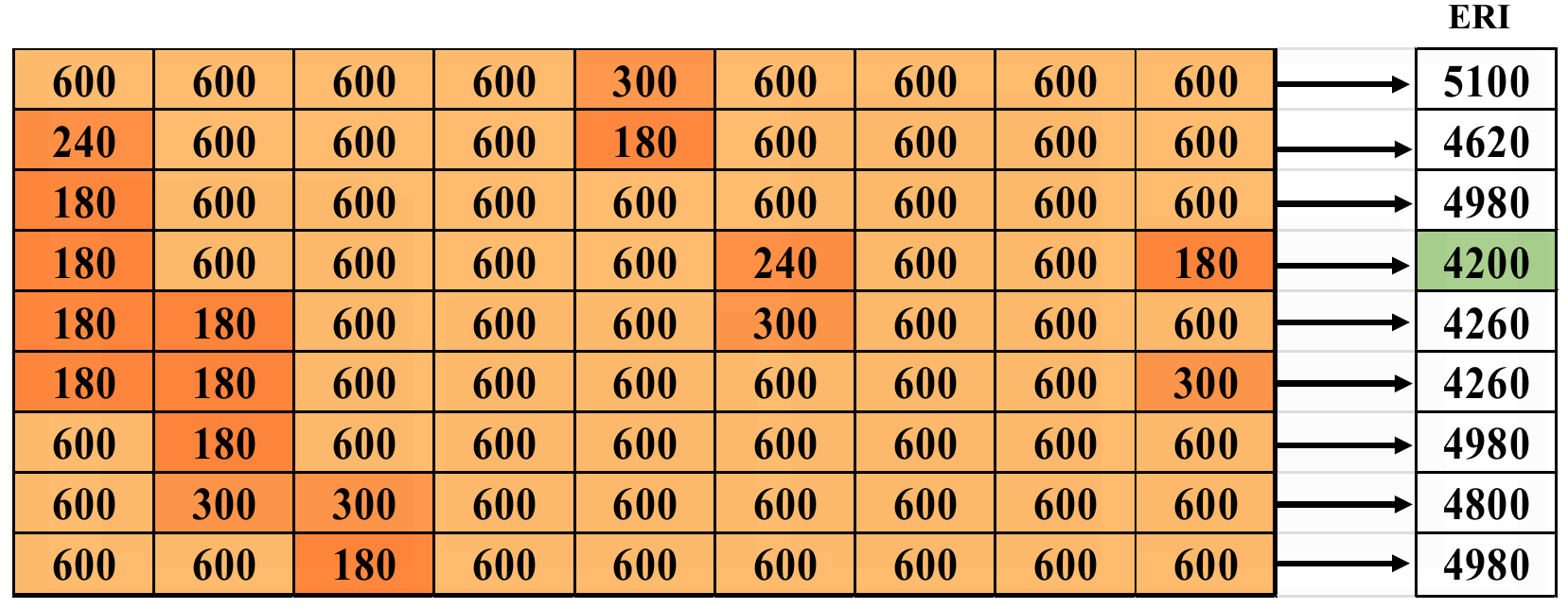}
		\caption{Optimal SuDoKu}
		\label{fig:fig:job_12_OPT_SUDOKU}	\end{subfigure}
    \begin{subfigure}[a]{0.36\textwidth}
		\includegraphics[width=\textwidth]{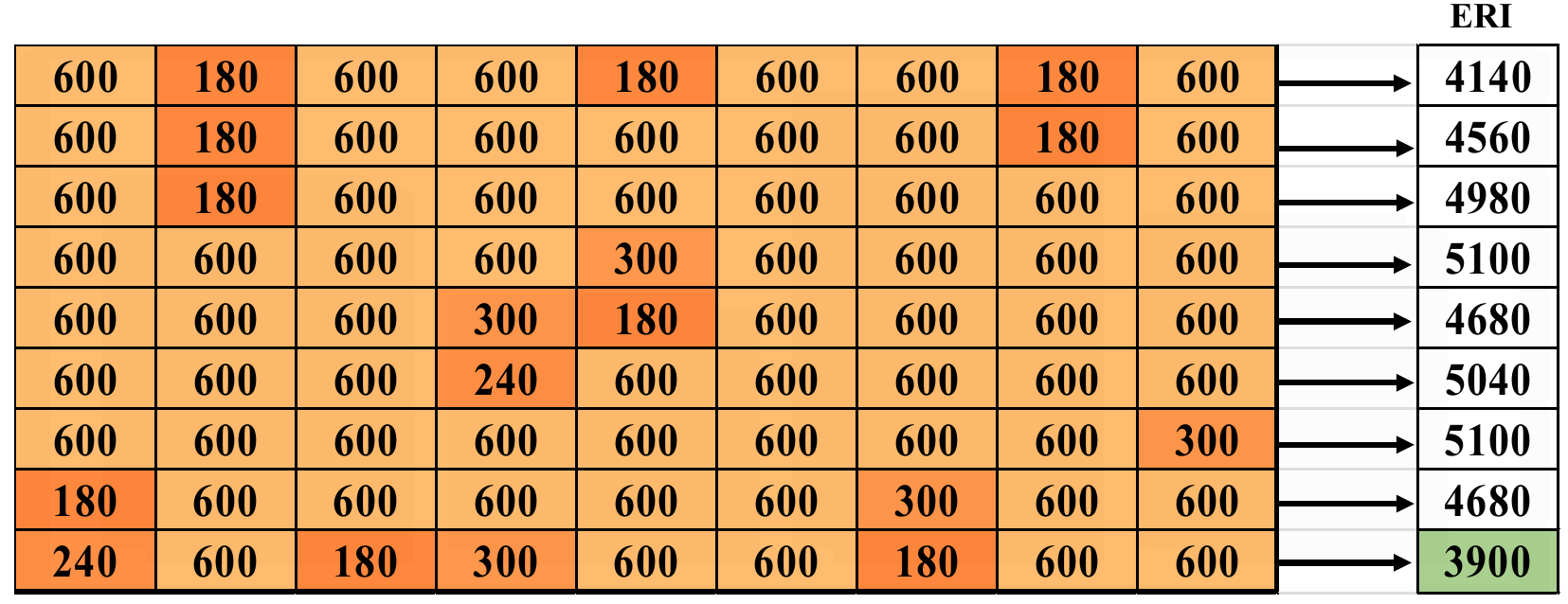}
		\caption{TTSA}
		\label{fig:job_12_TTSA} \end{subfigure}
    \begin{subfigure}[c]{0.36\textwidth}
		\includegraphics[width=\textwidth]{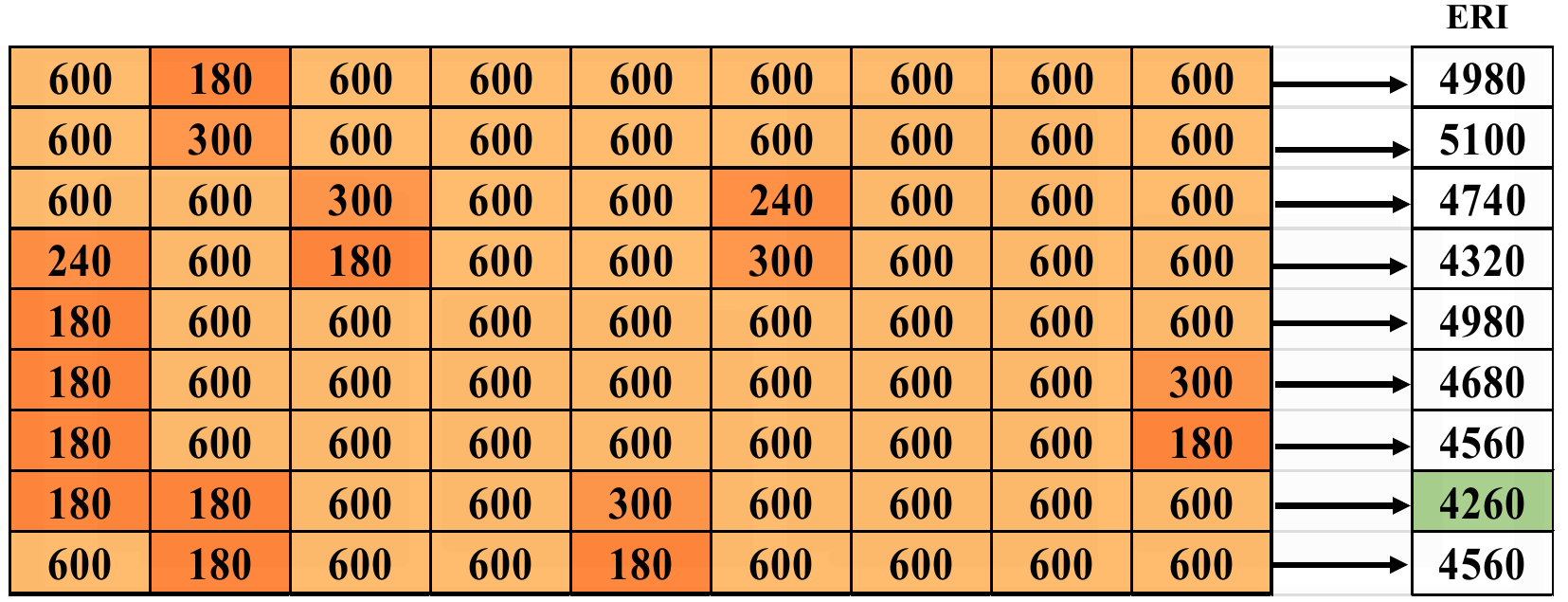}
		\caption{GMRA}
		\label{fig:fig:job_12_GMRA}	\end{subfigure}
    \begin{subfigure}[a]{0.36\textwidth}
		\includegraphics[width=\textwidth]{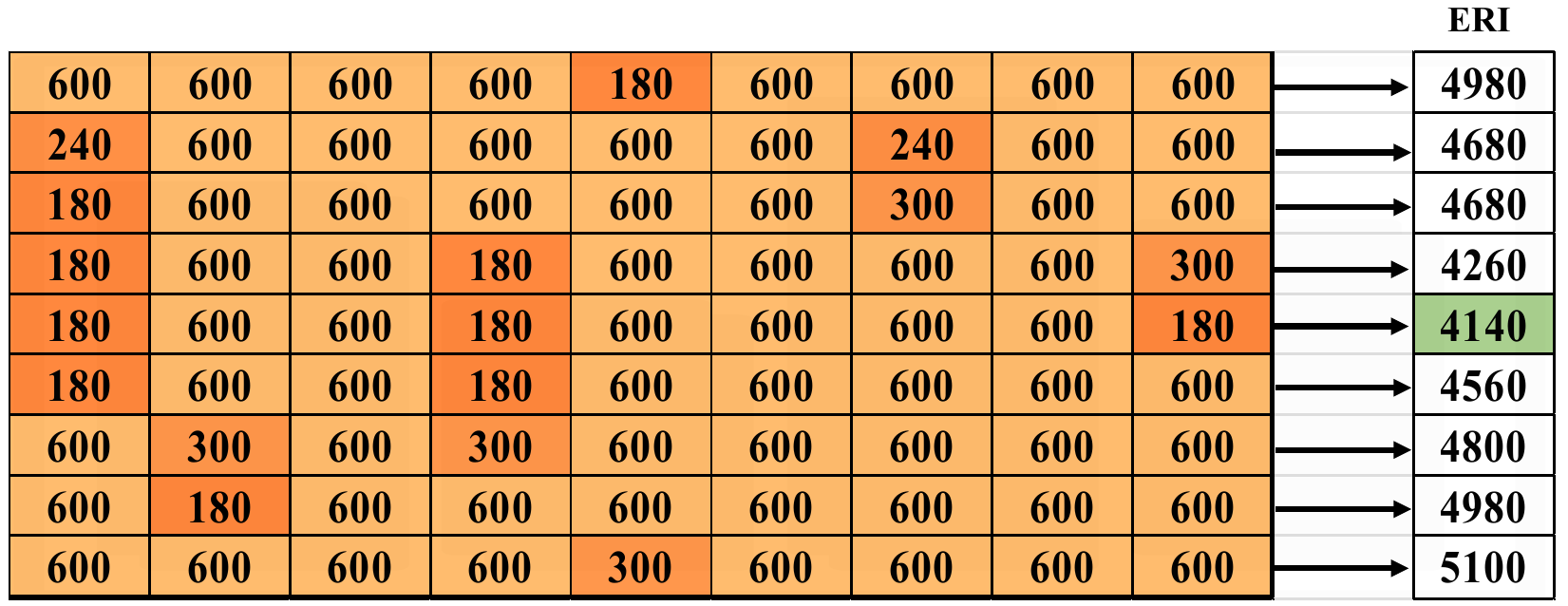}
		\caption{SPA}
		\label{fig:job_12_SPA} \end{subfigure}
    \begin{subfigure}[a]{0.36\textwidth}
		\includegraphics[width=\textwidth]{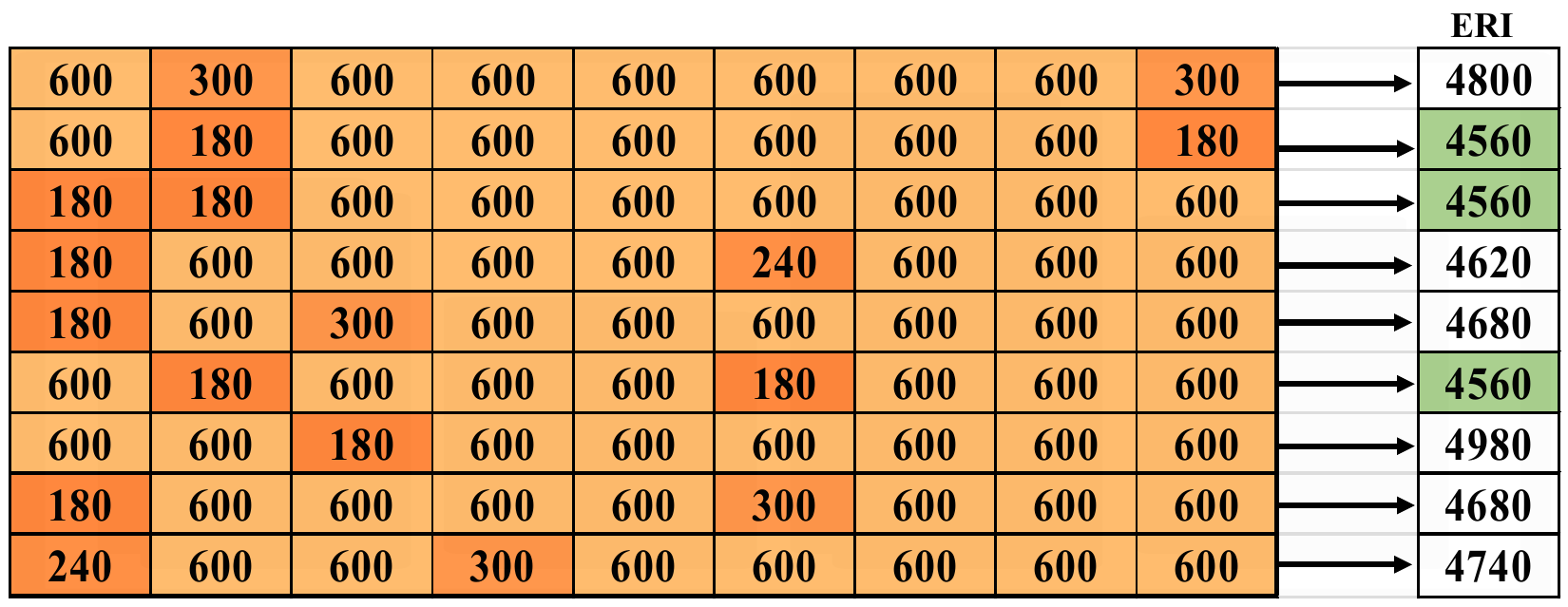}
		\caption{ILP}
		\label{fig:job_12_MILP} \end{subfigure}
    \caption{\textcolor{black}{Case-II:The shade dispersion of a $9\times 9$ SPV array at $14:30$ of the $349th$ day of year with (a) TCT, (b) SuDoKu \cite{sagar2020ku}, (c) Optimal SuDoKu \cite{krishna2019optimal}, (d) TTSA \cite{krishnan2022twisted}, (e) GMRA \cite{singh2023optimal}, (f) SPA \cite{cherukuri2021power}, and (g) Proposed ILP}}
    \label{fig:Shading_Patterns_28_10} \vspace{-4ex}
\end{figure}

\textcolor{black}{
In this case, $349th$ day of the year was considered for simulating the partial shading pattern. At $14:30$, $17$ PV modules were subjected to partial shading. The measured global irradiance was $600~W/m^2$, whereas the shaded modules experienced non-uniform irradiance levels of $300~W/m^2$, $240~W/m^2$, and $180~W/m^2$. The shading distribution corresponding to the conventional TCT configuration, together with the associated ERI expressed in ($W/m^2$), is presented in Fig.~\ref{fig:job_12_TCT}. The shade dispersion patterns and their respective ERI values obtained using the Sudoku \cite{sagar2020ku}, Optimal SuDoKu \cite{krishna2019optimal}, TTSA \cite{krishnan2022twisted}, GMRA \cite{singh2023optimal}, SPA \cite{cherukuri2021power}, and the proposed ILP-based SAR technique are illustrated in \figurename{~\ref{fig:fig:job_12_Sudoku}}, \figurename{~\ref{fig:fig:job_12_OPT_SUDOKU}}, \figurename{~\ref{fig:job_12_TTSA}}, \figurename{~\ref{fig:fig:job_12_GMRA}}, \figurename{~\ref{fig:job_12_SPA}}, and \figurename{~\ref{fig:job_12_MILP}}, respectively. A comparative analysis of these configurations indicated that the proposed ILP-based SAR approach attained the highest MERI of $4560~W/m^2$, thereby demonstrating its superior capability in mitigating the impact of partial shading.
}

\begin{figure}[t]
    \centering
	\includegraphics[width=0.48\textwidth]{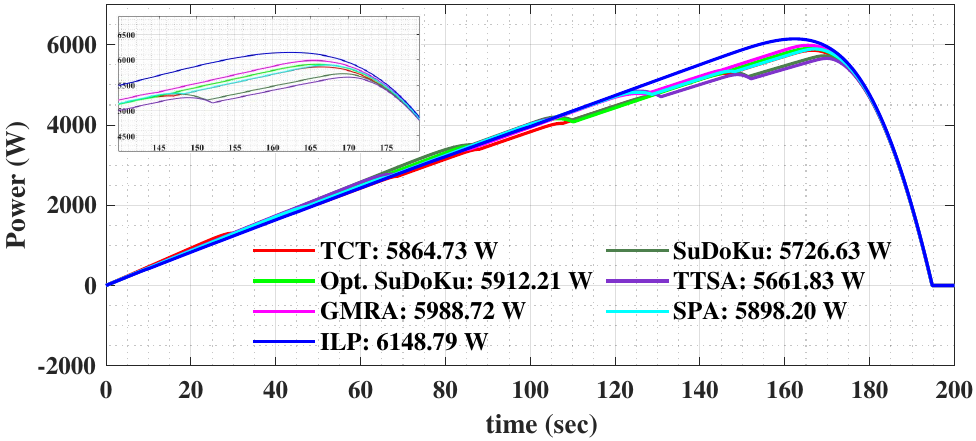}
	\caption{\textcolor{black}{Case-II: P-V characteristics for different reconfiguration strategies on the $349th$ day of year}}
    \label{fig:Case-II_P_V}
    \vspace{-3ex}
\end{figure}

\textcolor{black}{
The corresponding P-V characteristics of the PV array under SuDoKu, Optimal SuDoKu, TTSA, GMRA, SPA, and the proposed ILP-based SAR configurations are presented in \figurename{~\ref{fig:Case-II_P_V}}. It can be observed that the proposed method enables the extraction of the highest output power of $6148.79 ~W$ under the considered partial shading.
}

\begin{figure} [t]
    \centering
\begin{subfigure}[a]{0.24\textwidth}
	\includegraphics[width=\textwidth]{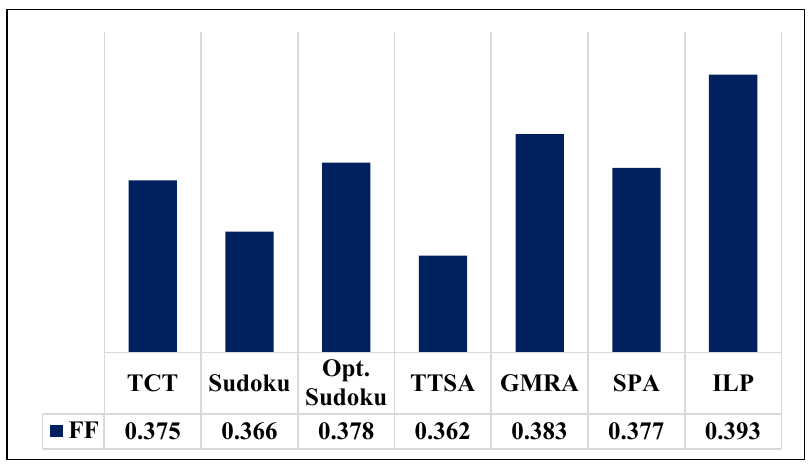}
	\caption{Fill Factor}
    \label{fig:Case-II_FF} \end{subfigure}
    \begin{subfigure}[a]{0.24\textwidth}
	\includegraphics[width=\textwidth]{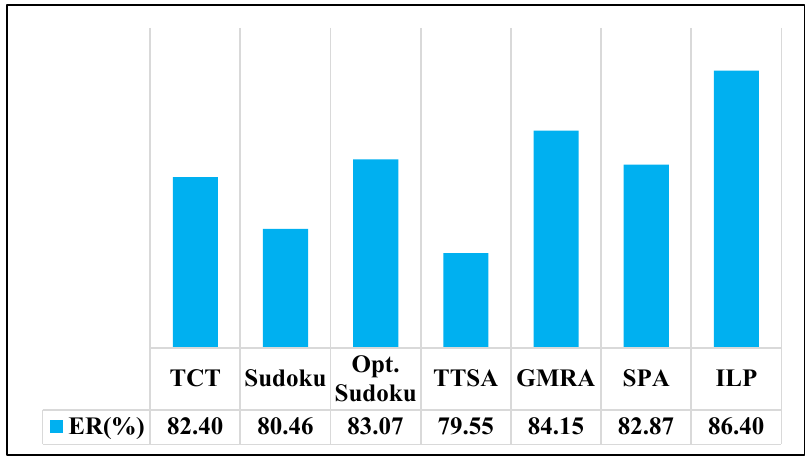}
	\caption{Execution Ratio}
    \label{fig:Case-II_ER} \end{subfigure}
    \begin{subfigure}[a]{0.24\textwidth}
    \includegraphics[width=\textwidth]{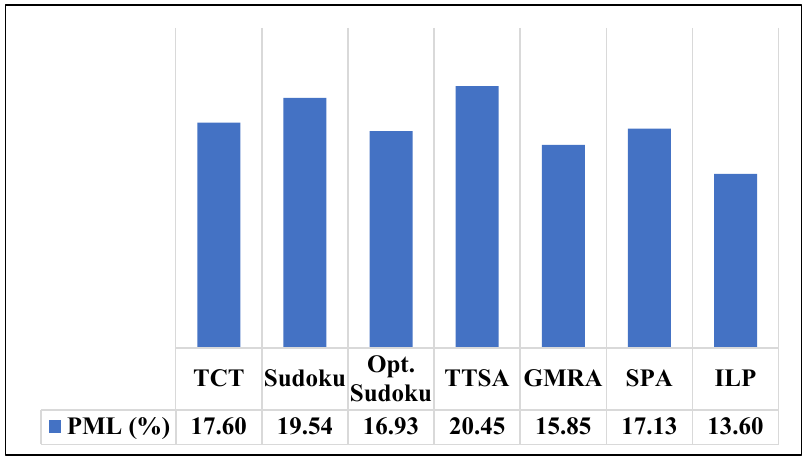}
	\caption{\% Mismatch Loss}
    \label{fig:Case-II_PML} \end{subfigure}
    \caption{Case-II: Performances indices of different array reconfiguration strategies on the $349th$ day of year} \label{fig:_PV_PI_II}
    \vspace{-3ex}
\end{figure}
\subsection{Comparison on performance matrices} \label{result_discussion_PM}
\textcolor{black}{
The detailed performance metrics for the two shading cases $166th$ and $349th$ days of year are shown in \figurename{~\ref{fig:_PV_PI}} and \figurename{~\ref{fig:_PV_PI_II}} respectively. For shading \textit{Case-I} (\figurename{~\ref{fig:_PV_PI}}), the proposed ILP based SAR technique outperforms all other methods across the evaluated indices. It achieves the highest FF of $0.696$ and the highest ER of $93.32\%$. In addition, the proposed method records the lowest PML of $6.68\%$, which is significantly lower than that of the conventional TCT configuration ($12.89\%$) as well as Sudoku ($7.78\%$), Optimal Sudoku ($8.81\%$), TTSA ($10.10\%$), GMRA ($12.53\%$), and SPA ($9.12\%$).
}
\textcolor{black}{
A similar trend observed for shading \textit{Case-II} shown in \figurename{~\ref{fig:_PV_PI_II}}. The ILP based technique achieves the highest FF of $0.393$ and the highest ER of $86.40\%$, while having the lowest PML of $13.60\%$ among all compared methods.}
\subsection{Comparison on energy generation} \label{energy_output}
\begin{figure}[ht]
    \centering
    \begin{subfigure}[a]{0.4\textwidth}
		\includegraphics[width=\textwidth]{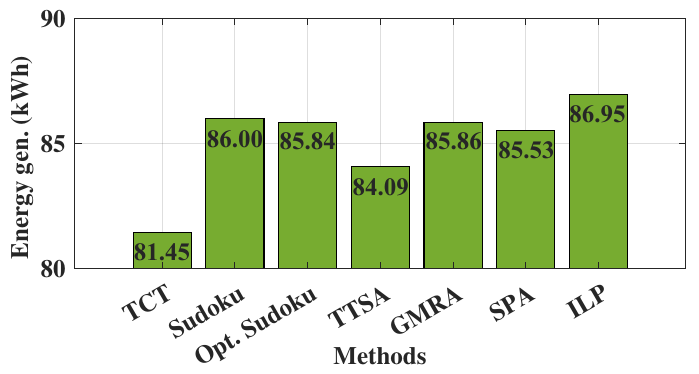}
		\caption{$166th$ Day of year}
		\label{fig:Energy_generation_29th_of_April} \end{subfigure}
    \begin{subfigure}[c]{0.4\textwidth}
		\includegraphics[width=\textwidth]{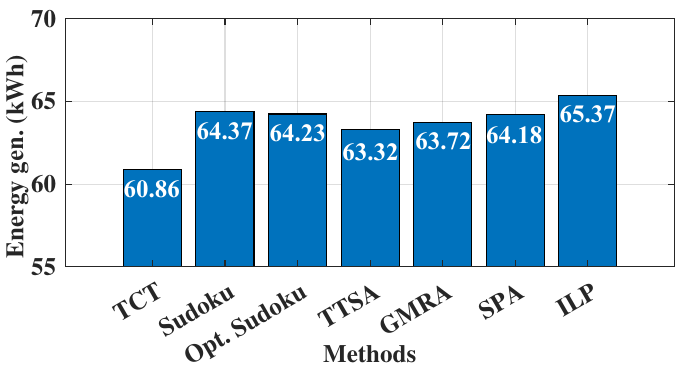}
		\caption{$349th$ Day of year}
		\label{fig:Energy_generation_28th_of_October}	\end{subfigure}
    \caption{\textcolor{black}{Total energy generation from a $9\times 9$ SPV array during clear sky days}}
    \label{fig:Effective_energy_generation} \vspace{-3ex}
\end{figure}

\textcolor{black}{This section evaluates daily energy generation under dynamic shading conditions. Shading variations were recorded on the $166th$ and $349th$ days of the year at half-hour intervals from $7:00$ to $16:00$. Observations identified $14$ distinct shading patterns on the $166th$ day (\figurename{~\ref{fig:Problem_formulation_29_04}}) and $18$ patterns on the $349th$ day (\figurename{~\ref{fig:Problem_formulation_28_10}}). The total daily energy generation using various reconfiguration methods was compared for both days (\figurename{~\ref{fig:Energy_generation_29th_of_April}} and \figurename{~\ref{fig:Energy_generation_28th_of_October}}). As shown in \figurename{~\ref{fig:Energy_generation_29th_of_April}}, the proposed method achieved a maximum daily energy output of $86.95~kWh$, outperforming alternative methods that yielded between $81.45~kWh$ and $86.0~kWh$. Similarly, \figurename{~\ref{fig:Energy_generation_28th_of_October}} demonstrates that the proposed method produced $65.37~kWh$, whereas the other techniques generated only $60.86~kWh$ to $64.37~kWh$. 
}

 

\section{Hardware Experiment}\label{lab:Hardware_exp}
\textcolor{black}{The effectiveness of the proposed ILP-based SAR technique was experimentally validated through hardware implementation. A small-scale laboratory prototype consisting of twelve $10~W_p$ PV modules arranged in a $4 \times 3$ configuration was developed to realize and test the proposed reconfiguration strategy. The detailed specifications of the PV modules used in the experimental setup are listed in \tablename{~\ref{tab: Module specification}}.
}

\subsubsection{Hardware Setup and Implementation Details} \label{lab:HW_exp_description}
\begin{figure}[t]
    \centering
	\includegraphics[width=0.4\textwidth]{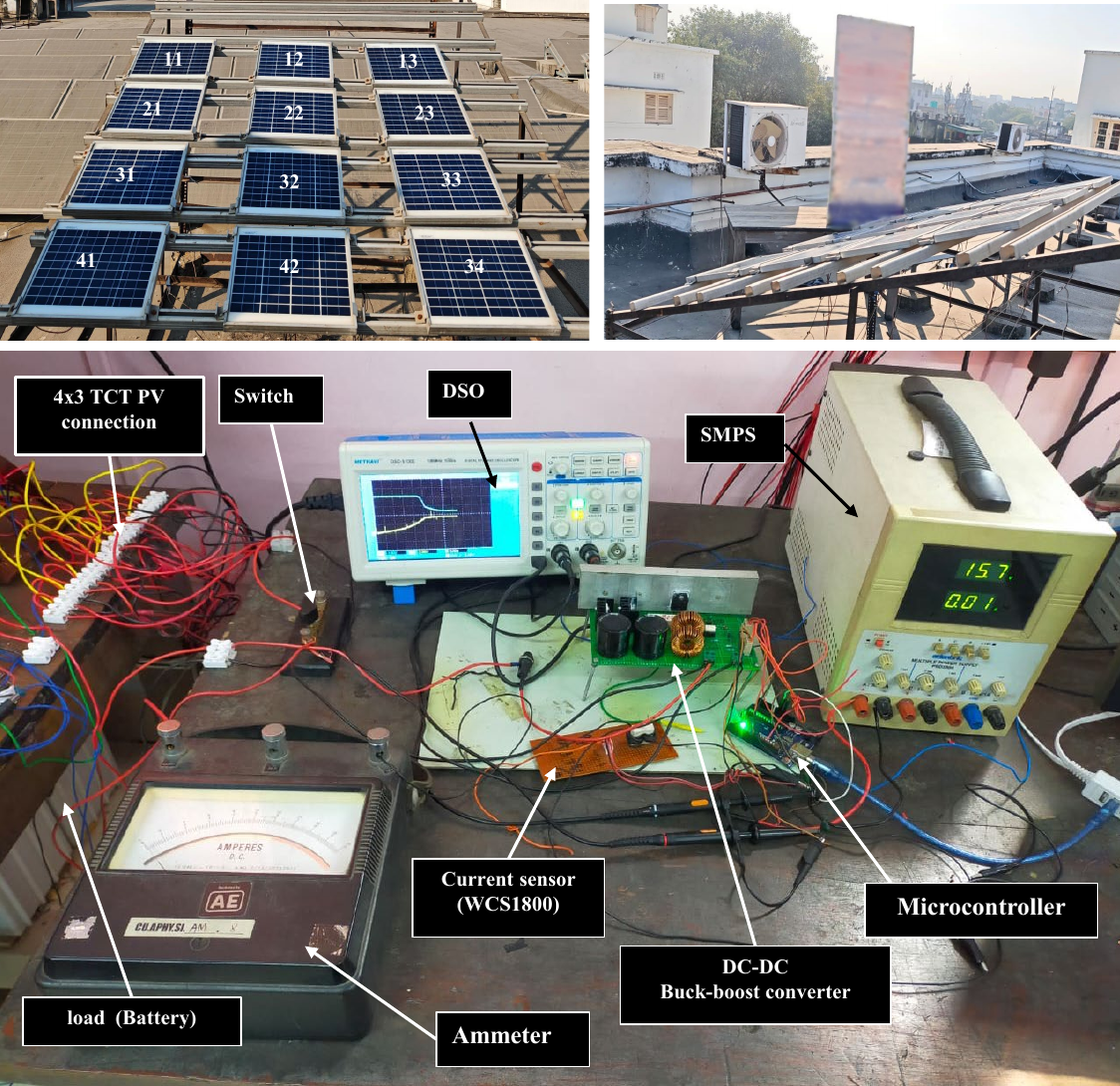}
	\caption{\textcolor{black}{Experimental hardware setup}}
    \label{fig:HW_setup}
    \vspace{-5ex}
\end{figure}

 \textcolor{black}{The complete hardware experimental setup is shown in \figurename{~\ref{fig:HW_setup}}. A $4\times 3$ SPV array is installed on the institute's rooftop ($22.57^\circ N, 88.36^\circ E$). As shown in \figurename{~\ref{fig:HW_setup}}, the PV array is connected to the load via a DC-DC buck-boost converter. In this setup, the load is a $24~V$ lead acid battery, which is charged using power from the solar PV array. The duty cycle of the buck-boost converter is managed by a microcontroller ($ATMEGA 2560$). To obtain the P–V characteristics of the PV array, the microcontroller is programmed to gradually reduce the duty cycle from $100\%$ to $0\%$, thus varying the array from short-circuit to open-circuit conditions. A $WCS1800$ Hall-effect current sensor was used to measure the array current. A resistive voltage divider was mounted on the converter board to measure the array voltage. The voltage and current signals were recorded using a digital storage oscilloscope (DSO).
 }

\begin{figure}[ht]
    \centering
    \begin{subfigure}[a]{0.32\textwidth}
		\includegraphics[width=\textwidth]{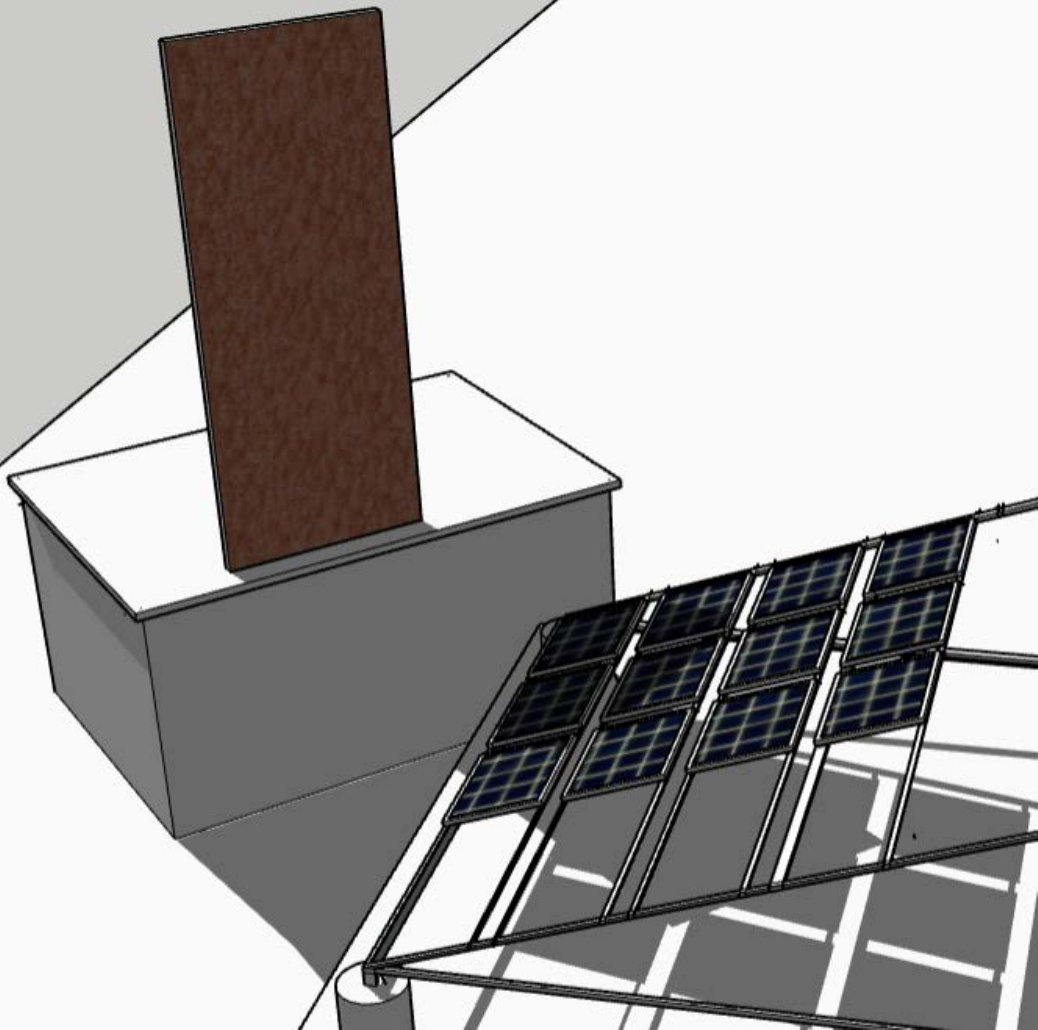}
		\label{fig:290_day_sv} \end{subfigure}
    \begin{subfigure}[c]{0.32\textwidth}
		\includegraphics[width=\textwidth]{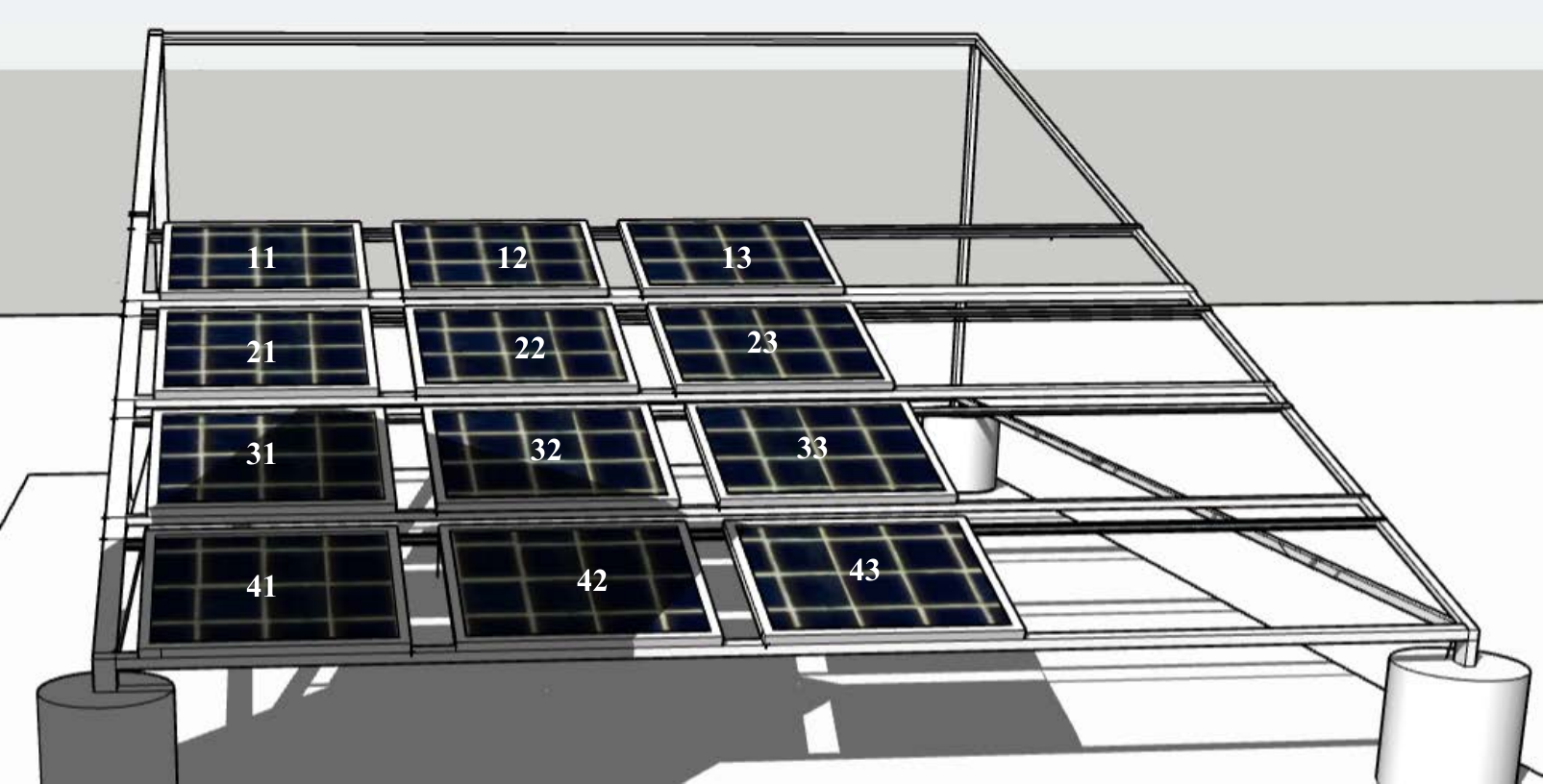}
		\label{fig:290_day_fv}	\end{subfigure}
    \caption{\textcolor{black}{Three-dimensional model of $4\times 3$ SPV array set up for laboratory experiment}}
    \label{fig:290_day_shading} \vspace{-3ex} 
\end{figure}

\begin{figure}[ht]
    \centering
    \begin{subfigure}[a]{0.15\textwidth}
		\includegraphics[width=\textwidth]{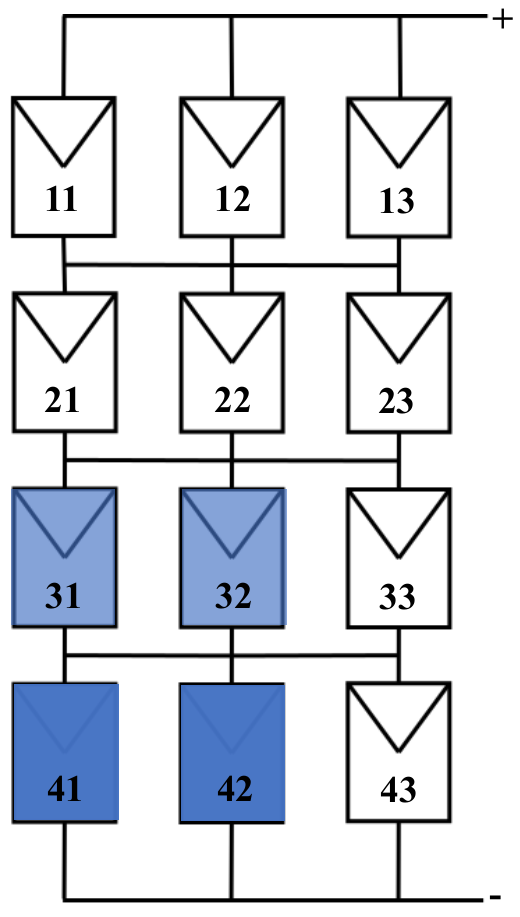}
		\caption{TCT matrix}
		\label{fig:TCT_config} \end{subfigure}
    \begin{subfigure}[a]{0.15\textwidth}
		\includegraphics[width=\textwidth]{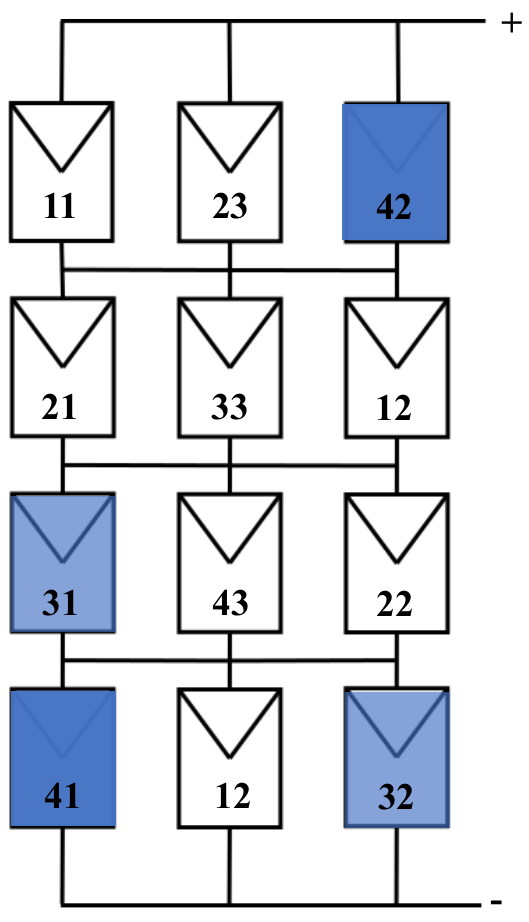}
		\caption{SPA matrix}
		\label{fig:SPA_config} \end{subfigure}
    \begin{subfigure}[a]{0.15\textwidth}
		\includegraphics[width=\textwidth]{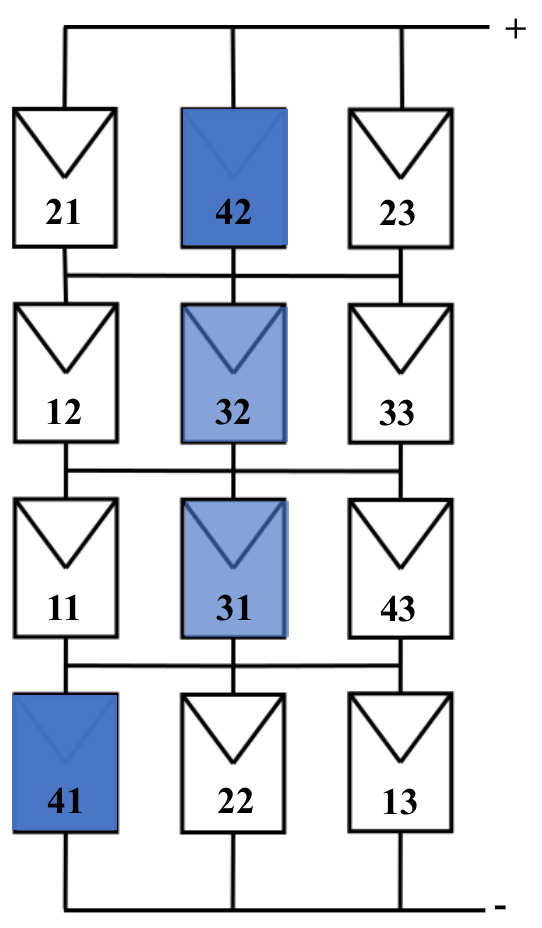}
		\caption{ILP matrix}
		\label{fig:MILP_config} \end{subfigure}
    \caption{\textcolor{black}{The shade dispersion of a $4\times 3$ SPV array for hardware experiment with (a) TCT, (b) SPA, and (c) Proposed ILP}}
    \label{fig:HW_matrix} \vspace{-2ex}
\end{figure}

\textcolor{black}{
An artificial partial shading condition was created using a rectangular placard ($5'\times 2'$) installed at a horizontal offset of $2.5'$ from the southwest corner of the PV array. The obstruction was oriented to produce consistent and repeatable shadow patterns on the array. Prior to experimentation, the complete setup was modeled in Google SketchUp as shown in \figurename{~\ref{fig:290_day_shading}} to replicate the array geometry and obstruction placement. The three-dimensional model facilitated visualization of shadow trajectories and preliminary evaluation of module-level shading under varying solar positions. The physical setup was implemented according to the design specifications, with shading profiles monitored at hourly intervals each month from October 15, 2024, to December 31, 2024. These observed patterns were compared with simulations generated using Google SketchUp for the corresponding periods. The physical observations matched the SketchUp-generated shading patterns exactly; consequently, the software was used to simulate shading profiles for the entire year. From this data, $30$ distinct shading patterns were identified, each characterized by a unique combination and number of shaded modules. These representative cases were then applied to a $4 \times 3$ PV array to determine a one-time optimal electrical configuration using the proposed Integer Linear Programming (ILP)-based reconfiguration strategy, as illustrated in \figurename{~\ref{fig:MILP_config}}. 
}\\

 \vspace{-2ex}
  \textcolor{black}{
\noindent \textbf{Shading Pattern for Hardware Evaluation:}\\
Among the thirty identified shading patterns, the case in which four modules are affected by shading was selected for experimental validation and comparison. In hardware evaluation the proposed method is compared with TCT and SPA method. The corresponding shade dispersion for the conventional TCT configuration, the SPA method and the proposed ILP-based reconfiguration are illustrated in \figurename{~\ref{fig:TCT_config}}, \figurename{~\ref{fig:SPA_config}}, and \figurename{~\ref{fig:MILP_config}}, respectively.
 }

\subsubsection{Experimental results} \label{lab:Discussion_HW}
\begin{figure}[ht]
    \centering
    \begin{subfigure}[c]{0.24\textwidth}
		\includegraphics[width=\textwidth]{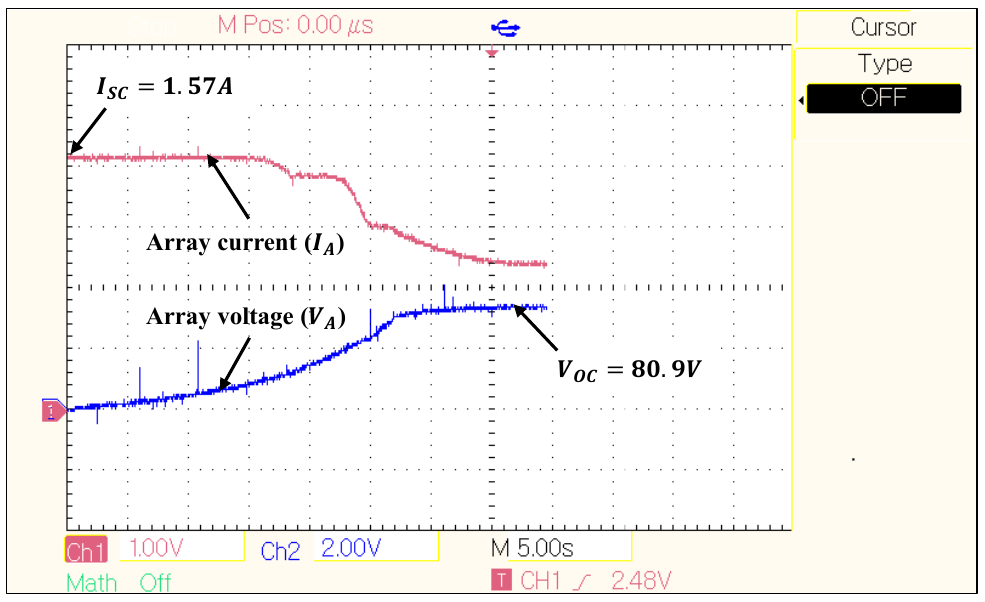}
		\caption{V-T, I-T characteristics of TCT}
		\label{fig:TCT_VI}	\end{subfigure}
    \begin{subfigure}[c]{0.24\textwidth}
		\includegraphics[width=\textwidth]{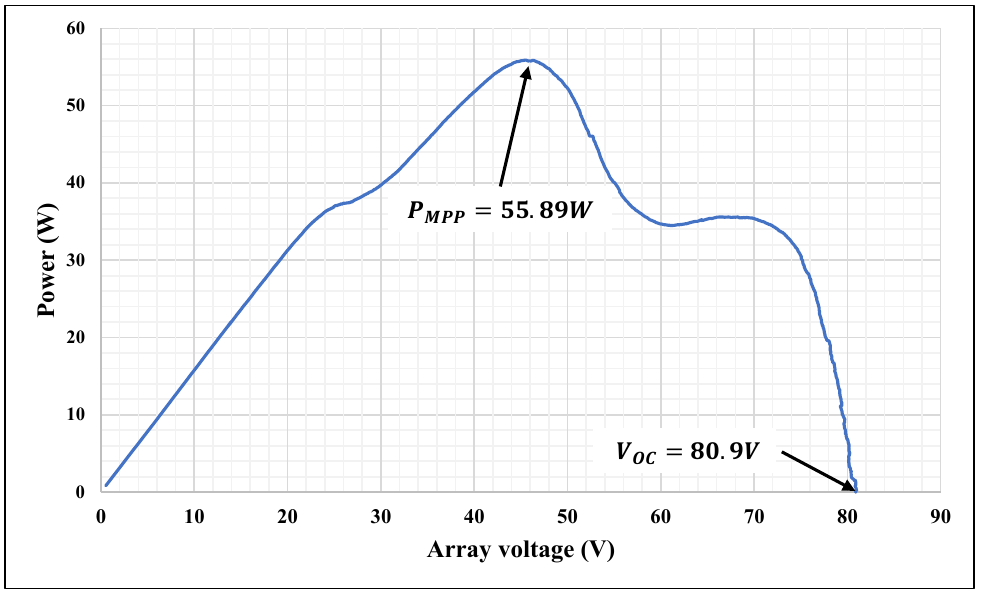}
		\caption{P-V characteristics of TCT}
		\label{fig:TCT_PV}	\end{subfigure}
    \begin{subfigure}[a]{0.24\textwidth}
		\includegraphics[width=\textwidth]{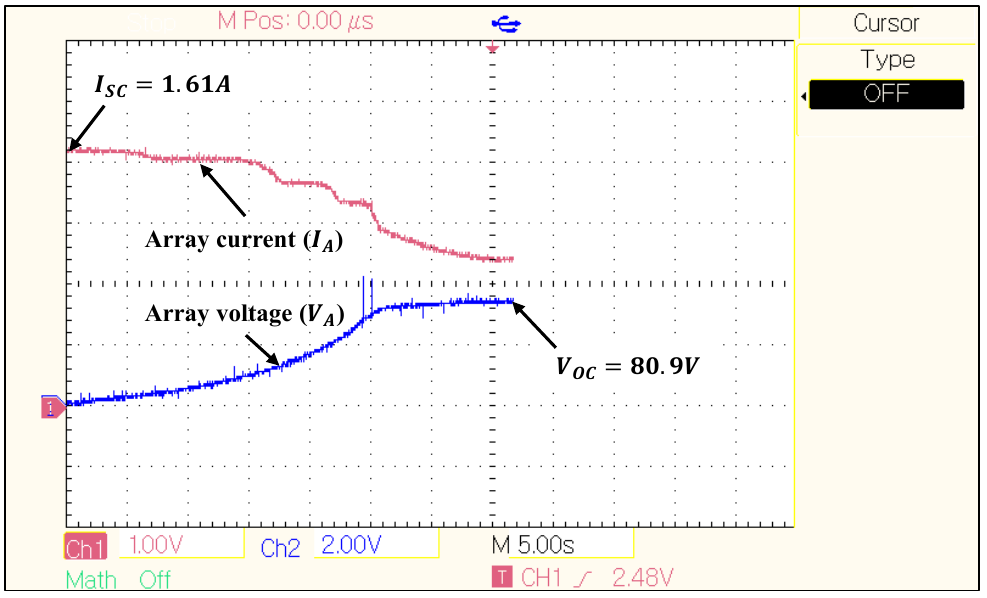}
		\caption{V-T, I-T characteristics of SPA}
		\label{fig:SPA_VI} \end{subfigure}
    \begin{subfigure}[c]{0.24\textwidth}
		\includegraphics[width=\textwidth]{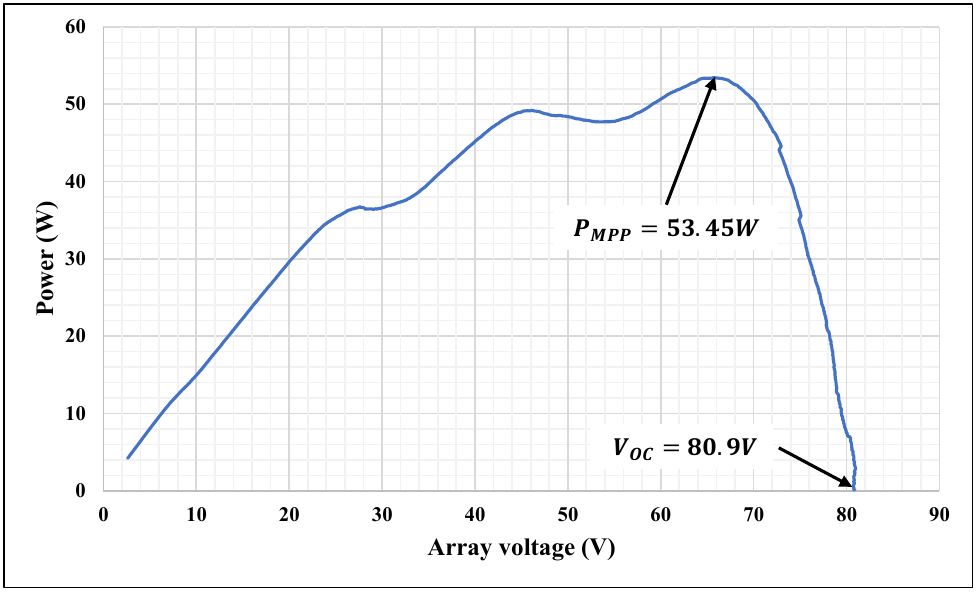}
		\caption{P-V characteristics of SPA}
		\label{fig:SPA_PV}	\end{subfigure}
    \begin{subfigure}[a]{0.24\textwidth}
		\includegraphics[width=\textwidth]{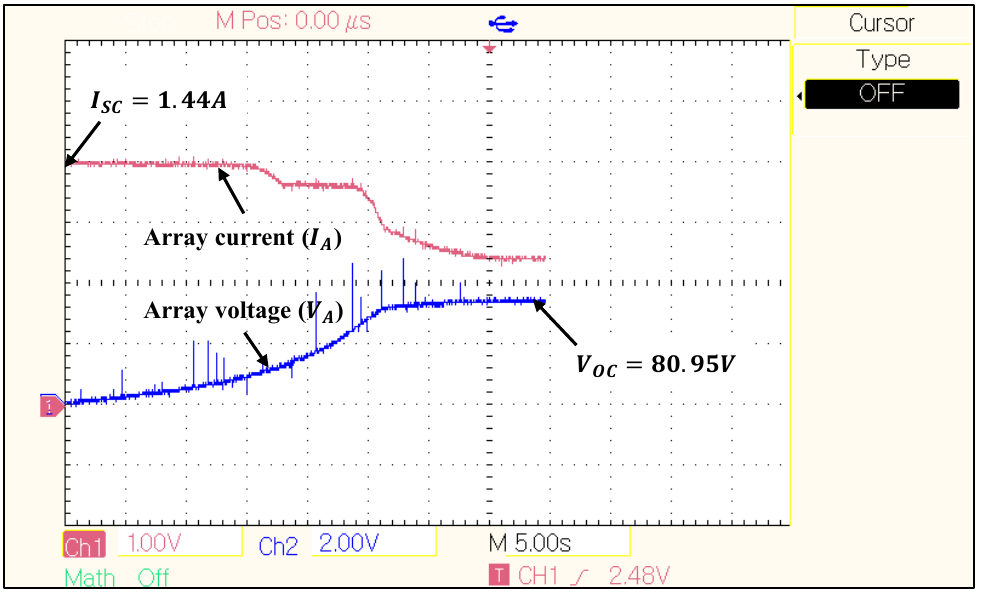}
		\caption{V-T, I-T characteristics of ILP}
		\label{fig:MILP_VI} \end{subfigure}
    \begin{subfigure}[c]{0.24\textwidth}
		\includegraphics[width=\textwidth]{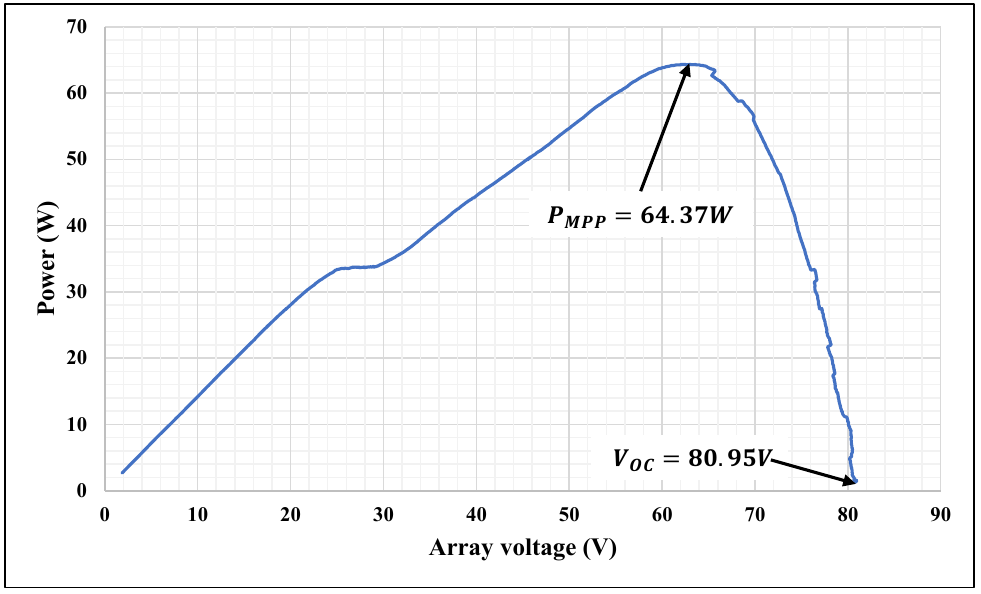}
		\caption{P-V characteristics of ILP}
		\label{fig:MILP_PV}	\end{subfigure}
    \caption{\textcolor{black}{$4\times3$ SPV array hardware experimental configuration and output results}}
    \label{fig:HW_exp_result} 
    \vspace{-3ex}
\end{figure}

\textcolor{black}{
The experiment was carried out $290th$ day of the year ($17th~October~2025$) at $14:00$ at an irradiance of $950~W/m^2$. The P–V characteristics of the PV array were obtained to determine the maximum extractable power from each configuration. \figurename{~\ref{fig:TCT_VI}} presents the oscilloscope waveform of the voltage and current variations, while \figurename{~\ref{fig:TCT_PV}} illustrates the corresponding P–V characteristics for the TCT configuration. Similarly, for the SPA and the proposed ILP-based reconfiguration, the oscilloscope waveforms of voltage and current are shown in \figurename{~\ref{fig:SPA_VI}} and \figurename{~\ref{fig:MILP_VI}}, respectively, and their corresponding P–V characteristics are depicted in \figurename{~\ref{fig:SPA_PV}} and \figurename{~\ref{fig:MILP_PV}}, respectively. It can be observed from \figurename{~\ref{fig:TCT_PV}}, \figurename{~\ref{fig:SPA_PV}}, and \figurename{~\ref{fig:MILP_PV}} that the proposed ILP-based reconfiguration technique achieves the highest maximum output power of $64.37~W$. In comparison, the TCT and SPA configurations deliver maximum powers of $55.89~W$ and $53.45~W$, respectively, thereby confirming the superior performance of the proposed method under the considered shading condition. 
}. 

\section{Conclusion} \label{sec:Conclussion}
Partial shading is a major challenge for the efficiency of solar photovoltaic systems, especially in urban environments. This study proposed an ILP based SAR method that accounts for shading caused by nearby fixed obstructions that varies throughout the day as well as across different seasons. Unlike existing SAR techniques that rely on assumed shading scenarios, the proposed approach utilizes actual shading data to determine an optimal, one-time module reallocation.

MATLAB simulations on a $9 \times 9$ TCT array demonstrate that the proposed approach consistently outperforms existing techniques. In Case-1, the proposed method achieved the lowest Percentage Mismatch Loss of $6.68\%$, compared to a range of $7.78\%$ to $12.53\%$ for alternative methods. Similarly, in Case-2, the proposed technique recorded a minimum PML of $13.6\%$, while other methods varied from \textcolor{black}{$15.85\%$ to $20.45\%$}. This optimized performance translated to the highest total daily energy generation of $86.95~kWh$ and $65.37~kWh$, respectively. 

Experimental validation on a $4 \times 3$ SPV prototype confirmed these simulation results. Under real world shading, the proposed configuration extracted a peak power of $64.37~W$, outperforming the TCT ($55.89~W$) and SPA ($53.45~W$) arrangements. Therefore, this ILP-based SAR provides a practical, computationally simple, and highly effective solution for maximizing energy yield in dynamic partial shading conditions.


\vspace{-2ex}
\section*{ACKNOWLEDGMENT}
This work was supported by the University of Calcutta under a minor innovative project grant [DPO/931 IRP dated 21/07/2025].

\bibliographystyle{IEEEtran}
\bibliography{Bibliography}
\vspace{-25ex}
\begin{IEEEbiography}[{\includegraphics[width=1in,height=1.25in,clip]{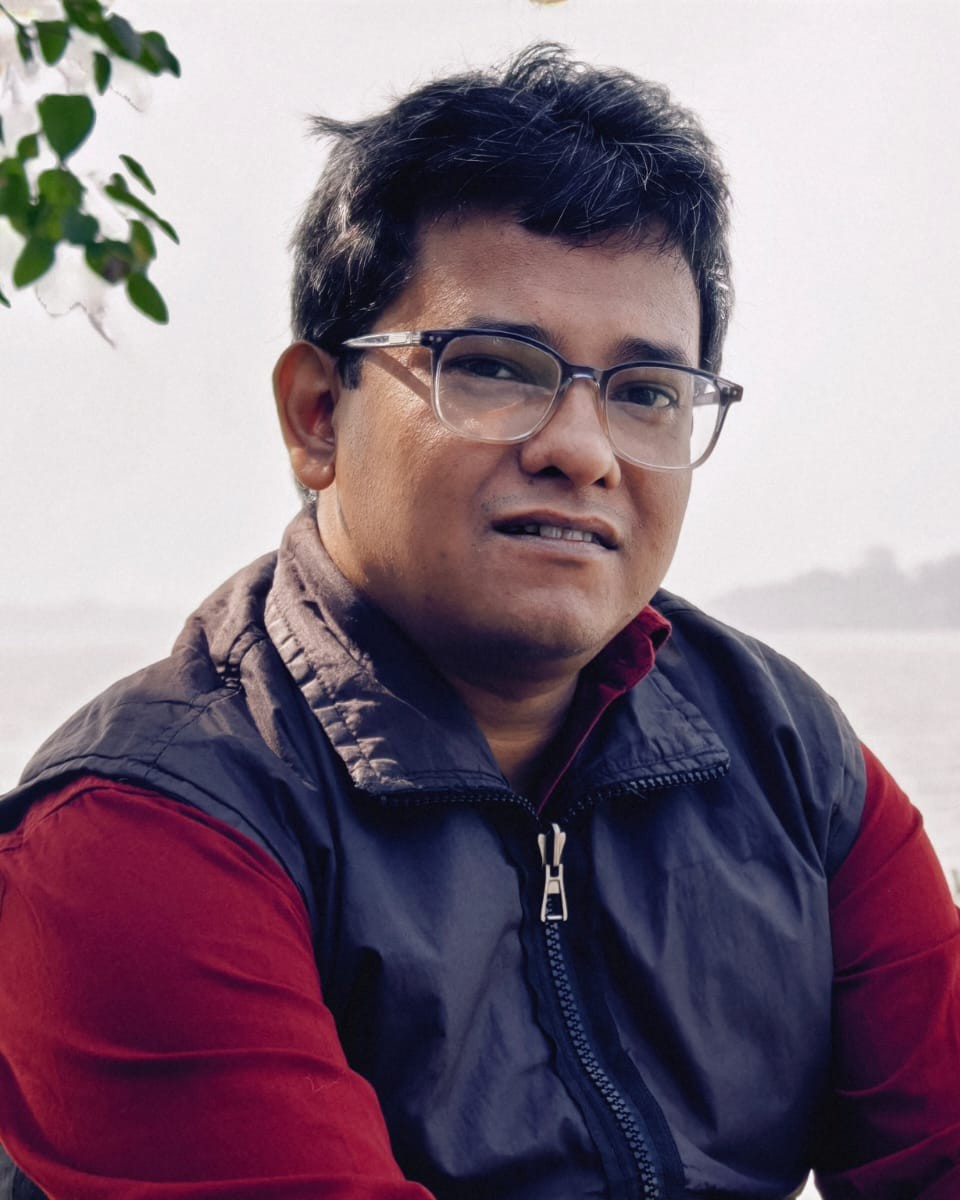}} ]
 {\textbf{Ushnik Chakrabarti}} received his M. Tech. degree in Power Systems (Electrical Engineering) from the Narula Institute of Technology, West Bengal, India. Following the completion of his master's degree, he served as a lecturer at various institutions. He is currently a Research Scholar in the Department of Applied Physics at the University of Calcutta, Kolkata, West Bengal, India. His research interests include fault detection and partial shading mitigation in solar photovoltaic systems. 
 \end{IEEEbiography}

\vspace{-25ex}
 \begin{IEEEbiography}[{\includegraphics[width=1in,height=1.25in,clip]{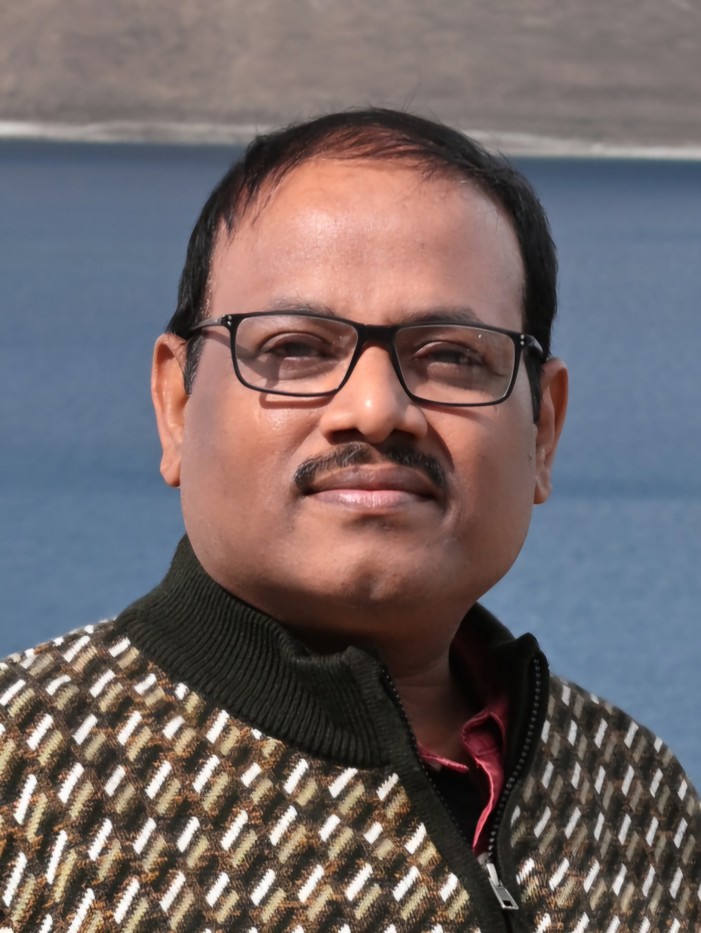}}]
 {\textbf{Binoy Kumar Karmakar}}
 (Senior Member, IEEE) received the M.Tech. degree in Power Systems and the Ph.D. degree from the Indian Institute of Technology Kharagpur, India. He is currently an Assistant Professor with the Department of Applied Physics, University of Calcutta, Kolkata, India. After completing his M.Tech., he worked with a startup company, where he contributed to the development of a power network simulator for research on electrical power systems. His current research interests include photovoltaic fault detection, partial shading mitigation, dynamic reconfiguration of PV arrays, and the integration of solar energy into smart grids.
 \end{IEEEbiography}

\end{document}